\documentclass[sigconf]{acmart}

\setcopyright{acmlicensed}
\copyrightyear{2026}
\acmYear{2026}
\setcopyright{cc}
\setcctype{by}
\acmConference[CCS '26]{Proceedings of the 2026 ACM SIGSAC Conference on Computer and Communications Security}{November 15--19, 2026}{The Hague, Netherlands}
\acmBooktitle{Proceedings of the 2026 ACM SIGSAC Conference on Computer and Communications Security (CCS '26), November 15--19, 2026, The Hague, Netherlands}
\acmDOI{10.1145/3830454.3846782}
\acmISBN{979-8-4007-2871-6/2026/11}

\usepackage[flushleft]{threeparttable}
\usepackage{multirow}

\usepackage{tikz}
\usepackage{amsmath}

\usepackage{filecontents}

\usepackage{amssymb}
\usepackage{xspace}
\usepackage{epsfig}
\usepackage{graphicx}
\usepackage{MnSymbol}
\usepackage{accents}
\usepackage{booktabs}
\usepackage{pifont}
\usepackage{url}
\usepackage{colortbl}
\usepackage{chngcntr}

\usepackage{algpseudocode}
\usepackage{paralist}
\usepackage{textcomp}
\usepackage{wasysym}
\usepackage{setspace}
\usepackage{textcomp}
\usepackage{soul}
\usepackage{diagbox}
\usepackage{pifont}
\usepackage{lipsum}
\usepackage[normalem]{ulem}
\usepackage{smartdiagram}
\usepackage{color}
\usepackage{framed}
\usepackage{enumerate}
\usepackage{tcolorbox}
\usepackage{fontawesome5}

\usepackage{subcaption}

\usepackage{soul}
\soulregister\cite7
\soulregister\ref7

\usepackage{float}
\usepackage{balance}
\usepackage{textcomp}

\usepackage{caption}
\usepackage{CJK}
\usepackage{microtype}

\usepackage{makecell}
\usepackage{booktabs}
\usepackage{graphicx}

\usepackage{pifont}

\usepackage{algorithm}
\usepackage{algpseudocode}

\usepackage{amsthm}

\usepackage{enumitem}
\setlist[itemize]{leftmargin=*, labelindent=0em}
\setlist[enumerate]{leftmargin=*, labelindent=0em, labelsep=1em}

\newcommand{\ignore}[1]{}

\newcommand{\hjs}[1]{#1}

\newcommand{\sssection}[1]{\noindent \textbf{#1. }}

\usepackage[misc]{ifsym}
\newcommand{\corr}{\textsuperscript{\Letter}}

\author{Jiasheng Huang}
\authornote{Both authors contributed equally to this work.}
\orcid{0009-0005-2645-0100}
\affiliation{%
  \institution{Tsinghua University}
  \country{Beijing, China}}
\email{huangjc21@mails.tsinghua.edu.cn}

\author{Mingxuan Liu}
\authornotemark[1]
\orcid{0000-0002-2163-6505}
\affiliation{%
  \institution{Zhongguancun Laboratory}
  \country{Beijing, China}}
\email{liumx@mail.zgclab.edu.cn}

\author{Pei Chen}
\orcid{0009-0001-8088-316X}
\affiliation{%
  \institution{Fudan University}
  \country{Shanghai, China}}
\email{peichen19@fudan.edu.cn}

\author{Baojun Liu\corr}
\authornote{\Letter\ Corresponding authors.}
\orcid{0000-0002-9032-8063}
\affiliation{%
  \institution{Tsinghua University}
  \city{Beijing}
  \country{China}}
\email{lbj@tsinghua.edu.cn}

\author{Yiming Zhang}
\orcid{0000-0002-6774-5299}
\affiliation{%
  \institution{Tsinghua University}
  \country{Beijing, China}}
\email{zhangyiming@tsinghua.edu.cn}

\author{Geng Hong}
\orcid{0000-0003-1811-9432}
\affiliation{%
  \institution{Fudan University}
  \country{Shanghai, China}}
\email{ghong@fudan.edu.cn}

\author{Zhenrui Zhang}
\orcid{0009-0003-9826-0141}
\affiliation{%
  \institution{Baidu Inc.}
  \country{Beijing, China}}
\email{zhangzhenrui@baidu.com}

\author{Hai Yang}
\orcid{0009-0008-7814-1862}
\affiliation{%
  \institution{Baidu Inc.}
  \country{Beijing, China}}
\email{yanghai01@baidu.com}

\author{Haixin Duan}
\orcid{0000-0003-0083-733X}
\authornote{Also with Tsinghua University.}
\affiliation{%
  \institution{Quancheng Laboratory}
  \country{Jinan, China}}
\email{duanhx@tsinghua.edu.cn}

\author{Hui Jiang\corr}
\authornotemark[2]
\orcid{0009-0005-0306-6691}
\affiliation{%
  \institution{Tsinghua University}
  \city{Beijing}
  \country{China}}
\additionalaffiliation{%
  \institution{Baidu Inc.}
  \city{Beijing}
  \country{China}}
\email{jianghui01@baidu.com}
\renewcommand{\shortauthors}{Huang et al.}

\begin{document}
\title{One Click to Leak: Characterizing the Real-World Usage and Threat Impact of MNO-based Single Sign-On Websites}





\begin{abstract}
%
Mobile Network Operator (MNO)-based Single Sign-On (MSSO) is a password-free authentication framework that relies on mobile data sessions. Unlike traditional SSO, it shifts the Identity Provider (IdP) to the MNO and the authentication anchor to the Service Provider (SP). MSSO is increasingly deployed and has expanded from mobile apps to websites. However, the emerging ecosystem and security risks of web-based MSSO remain largely unexplored.


To fill this gap, we first analyze mainstream MSSO deployments and identify a 3-phase workflow with three trust defects enabling the trust hijacking threat. We further demonstrate \textit{One-Click-to-Leak (OCL)} attacks, where a single webpage visit can leak sensitive identity information (e.g., phone numbers).
To systematically assess real-world deployment and risk, we conduct the first large-scale, longitudinal study of web-based MSSO with a leading security company. We design a hierarchical detection framework leveraging passive DNS correlations and URL reconstruction from search data to identify MSSO-enabled websites. Over one year, we identified 116,852 such website URLs across 729 apex domains.
\hjs{Unfortunately, 73.6\% of these URLs exhibit at least one trust defect: 69.4\% expose developer credentials, and 27.1\% issue high-privilege tokens before user consent, showing that OCL-enabling trust defects are widespread in real deployments. Structurally, 31.8\% of the 729 apex domains rely on Resellers, obscuring the downstream SP from the MNO in the analyzed flows. By analyzing scripts on MSSO-enabled sites, we identify 101 websites strongly associated with OCL attack behavior. In collaboration with our partner, we trace a representative upstream platform that was subsequently seized by law enforcement and uncover a monetized underground ecosystem. Sanitized backend data shows that it collected 14,100 users' phone numbers within three days and linked them to sensitive information such as browsing activity.}
\ignore{
Unfortunately, xx websites exhibit trust defects: 386 expose developer credentials, and xx fail to validate request origins, making them vulnerable to OCL attacks. By analyzing scripts on MSSO-enabled sites, we identify 101 websites likely conducting OCL attacks. In collaboration with our partner, we take over the infrastructure of one such site and uncover a monetized underground ecosystem. Sanitized backend data shows that it collected 14,100 users’ phone numbers within three days and linked them to sensitive information such as browsing activity.
}
\hjs{Our work provides} a comprehensive study of the real-world deployment landscape and security implications of web-based MSSO. Through responsible disclosure, our work helps secure the mobile authentication ecosystem.

\end{abstract}

\maketitle


\section{Introduction}
\label{sec:intro}
\begin{figure}[t]
    \centering
    \includegraphics[width=1.0\linewidth]{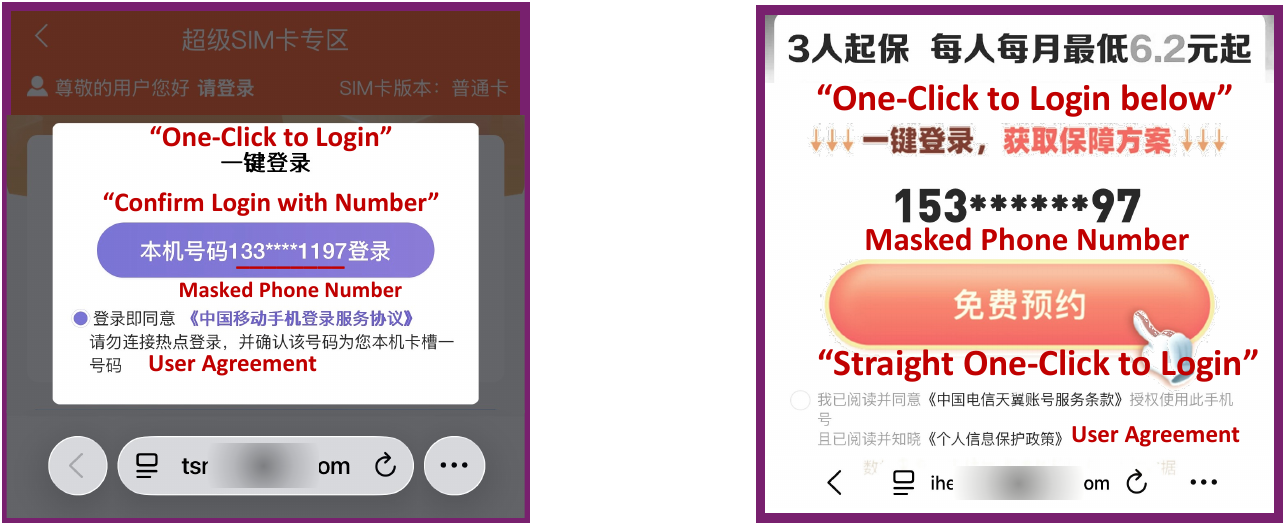}
    \caption{\hjs{Examples of Web-Based MSSO.}}
    \label{fig:web-msso-demo}
\end{figure}

Mobile Network Operator (MNO)-based Single Sign-On (MSSO) has emerged as a password-free authentication framework that uses a user's active mobile data session (e.g., 4G/5G) as an identity signal, \hjs{also termed one-click login}.
In an MSSO flow, the MNO acts as the Identity Provider (IdP): it maps the user's mobile data session to a verified subscriber identity, typically a phone number, and allows a Service Provider (SP) to authenticate the user through a one-click login flow, as shown in Figure~\ref{fig:web-msso-demo}. 
\hjs{A similar delegation pattern appears in Web SSO, although its trust anchor and identity source differ: an independent Web IdP authenticates an IdP account and asserts identity to the SP through browser redirects.}
By reducing password entry, manual SMS OTP retrieval, and the browser redirects required by conventional Web SSO, MSSO offers a low-friction authentication experience for login, onboarding, and account verification. 

\hjs{Originally deployed in native mobile applications, MSSO is now expanding onto the open web, allowing ordinary webpages to initiate this one-click authentication flow. 
Within MSSO, this delegated relationship is implemented differently across client environments: native applications encapsulate the authentication logic in proprietary SDKs and compiled binaries, whereas web deployments execute it as JavaScript in the browser, where authentication parameters, tokens, and developer credentials may appear in readable code. 
However, this expansion exposes MSSO's delegated trust model to the open architecture of the web. 
Prior work has examined account and session security in Web SSO and weaknesses in native one-tap authentication~\cite{GhasemisharifRC18,DBLP:conf/dsn/ZhouHCNLG22,DBLP:journals/tifs/CuiCFB23}, but it does not fully characterize how MSSO's delegated trust boundaries are established and enforced when authentication logic runs in the browser.}

\noindent \textbf{Empirical Study: Three-phase Workflow.}
To systematically investigate web-based MSSO, we first reconstruct the real-world deployment and operation of it via empirical analysis.
By analyzing provider documentation and real-world deployments, we confirm 14 MSSO Providers that typically offer services through APIs or JavaScript SDKs, and identify two roles: direct MNO providers, which perform network-level identity verification, and Third-Party Resellers, which rebundle MNO authentication capabilities through unified developer-facing interfaces. We further reconstruct a common three-phase workflow of web-based MSSO: \textit{Pre-authentication}, \textit{Token Issuance}, and \textit{Backend Verification}. This workflow captures how the delegated trust chain operates on the web and provides the basis for analyzing where it breaks.

\noindent \textbf{Trust Hijack Threat.}
\hjs{Furthermore, we uncover three \textit{Trust Defects} in this delegated trust chain.}
\textit{Trust Defect 1: Unenforced User Consent} arises because the MNO cannot verify at the protocol level that the user approved an authentication request. \textit{Trust Defect 2: Static Credential Exposure} stems from the use of static SP developer credentials in readable web assets, where exposure undermines the credential-based boundary for authenticating legitimate SPs. \textit{Trust Defect 3: Origin Validation Blindspot} results from the MNO's inability to reliably identify the true request origin, especially when Third-Party Resellers mediate authentication. Together, these defects break the delegated trust chain and enable a broader \textit{Trust Hijack Threat}, in which an adversary hijacks the trust relationships embedded in MSSO. Its concrete attack form is \textit{One-Click-to-Leak (OCL)}: a single webpage visit over a mobile data session can leak the user's verified phone number without an explicit login action.

\noindent \textbf{Longitudinal Real-World Measurement.}
Measuring this threat at web scale is challenging. MSSO integrations are sparse and distributed, and OCL-relevant behavior appears in authentication traffic and JavaScript execution rather than static page content alone. Brute-force crawling is computationally inefficient, while monitoring real users' authentication traffic would be costly and privacy-invasive. 
Therefore, we collaborated with a leading search engine company, Baidu, to design and implement OCL Detector, a scalable measurement framework with a staged pipeline:
it uses passive DNS correlation to narrow the broader web to domains associated with MSSO authentication, maps these domains to candidate URLs using search-engine data, verifies MSSO integrations through targeted dynamic probing in a simulated mobile environment, and applies script-level abuse analysis to identify high-risk MSSO usage.

Using OCL Detector, we conduct a one-year measurement of web-based MSSO deployments. We identify 116,852 MSSO-enabled URLs across 729 apex domains, spanning high-trust sectors such as finance, e-commerce, and travel. \textit{The results show that Trust Defects are widespread: 73.6\% of these URLs exhibit at least one Trust Defect, 69.4\% expose developer credentials, and 27.1\% issue high-privilege tokens before user consent.} At the domain level, 31.8\% of the 729 apex domains rely on Third-Party Resellers, leaving MNOs structurally blind to the true request origin. These findings show that Trust Defects are prevalent across the web-based MSSO ecosystem rather than isolated deployment mistakes.

Beyond prevalence, we show that OCL is already exploited in the wild. By analyzing scripts that trigger MSSO calls, we identify 101 websites strongly associated with abusive MSSO exploitation through scripts supplied by 5 upstream platforms. Through collaboration with an industry partner, our findings supported a law-enforcement investigation that led to the seizure of several abusive platforms. We analyze sanitized backend code and data from one representative seized platform. 
The seized platform collected 14,100 unique phone numbers within three days and linked these identifiers to sensitive web context such as browsing activity and search keywords, revealing a monetized pipeline for large-scale user de-anonymization.


\noindent \textbf{Contribution.} This paper makes the following contributions:

\noindent $\bullet$ \textit{Web-based MSSO Characterization:} 
Through systematic empirical study, we characterize the provider landscape, delegated trust model, and common three-phase workflow of MSSO. 

\noindent $\bullet$ \textit{Novel Trust Hijack Threat:}
We identify three Trust Defects in web-based MSSO deployments that jointly enable the \textit{Trust Hijack Threat}. We further identify \textit{One-Click-to-Leak (OCL)} as a concrete attack, leaking users' verified phone numbers through simple mobile-data webpage access without explicit login.

\noindent $\bullet$ \textit{Large-Scale Measurement:}
We design OCL Detector, a scalable measurement framework. Combining passive DNS correlation and search engine perspectives, we conduct large-scale measurements on real-world MSSO deployments and underlying trust defects. Over one year, we identify 116,852 MSSO-enabled URLs across 729 apex domains, and quantify that 73.6\% of web-based MSSO deployments exhibit at least one trust defect. 
We further pinpoint 101 abuse-related websites linked to OCL exploitation via five upstream platforms. Analysis of anonymized backend data from a representative takedown service revealed 14,100 collected phone numbers within three days, highlighting the significant risks of OCL.

\noindent $\bullet$ \textit{Defense Insight and Disclosure:}
\hjs{To mitigate the identified risks, we propose practical defense mechanisms. We also conduct responsible disclosure to affected MNOs, SPs, and law-enforcement agencies and provide active assistance with our collaborators. Among the domains identified by OCL Detector, we observed that at least 148 affected sites had deployed recommended defenses, including safer MSSO implementations such as masked-number completion.}

\section{MNO-based SSO Preliminaries}
\label{sec:background}
In this section, starting by illustrating the evolution of web authentication, we introduce the Mobile Network Operator (MNO)-based single sign-on (MSSO) and its landscape.

\begin{figure*}[t]
    \centering
    \includegraphics[width=1.0\linewidth]{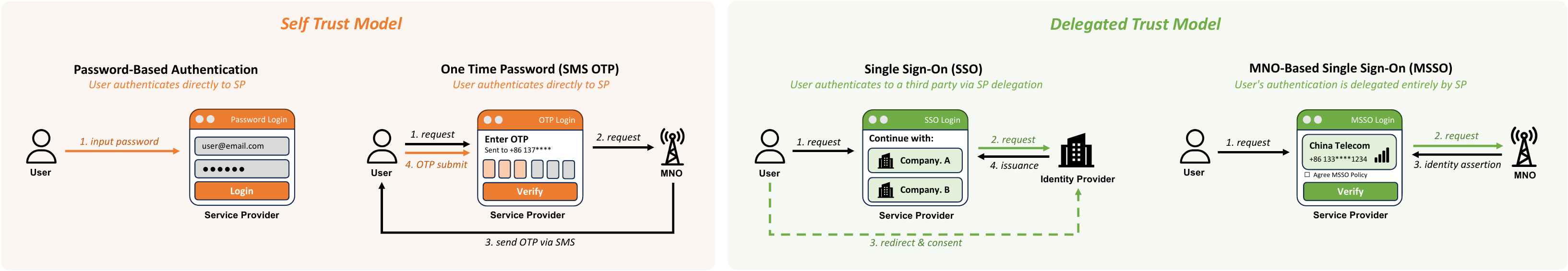}
    \caption{Four Different Types of Authentication}
    \label{fig:turst-models}
\end{figure*}

\subsection{Evolution of Web Authentication}
\label{sec2:evolution}
User authentication serves as the critical line of defense for account security. To better balance security and usability in the authentication process, mechanisms have evolved from peer-to-peer schemes (e.g., password-based login) toward trust-delegated authentication frameworks (e.g., Web Single Sign-On).
Figure~\ref{fig:turst-models} summarizes this progression through four representative authentication schemes.

Password-based authentication treats the user as the sole trust anchor, requiring the user to memorize and input a pre-negotiated secret into the login interfaces of the Service Provider (SP)~\cite{DBLP:journals/cacm/BonneauHOS15, DBLP:conf/sp/BonneauHOS12}. 
To reduce the memorization overhead of passwords, SMS One-Time Passwords (OTP) introduce an MNO-assisted verification step~\cite{DBLP:journals/rfc/rfc4226, DBLP:journals/rfc/rfc6238, DBLP:conf/acsac/MaFLLNOBDMJ19}.
This mechanism relies on MNOs to verify the user's identity via phone number--subscriber binding and issue OTPs as login credentials.
\ignore{
To reduce the memorization overhead of passwords, SMS One-Time Passwords (OTP) introduced an initial notion of delegation~\cite{DBLP:journals/rfc/rfc4226, DBLP:journals/rfc/rfc6238, DBLP:conf/acsac/MaFLLNOBDMJ19}. 
为了减少记忆密码的开销，短信一次性密码 (OTP) 引入了委托的初始概念 ~ \cite { DBLP:journals/rfc/rfc4226, DBLP:journals/rfc/rfc6238, DBLP:conf/acsac/MaFLLNOBDMJ19 } 。
This mechanism shifts trust from users to Mobile Network Operators (MNOs), which verify identity via phone number–subscriber binding and issue OTPs for login.
}
To improve usability, the web ecosystem adopts Web Single Sign-On (SSO) (e.g., OAuth 2.0)~\cite{rfc6749, DBLP:conf/eurosp/MainkaMSW17, DBLP:conf/ccs/SunB12, DBLP:journals/csur/AlacaO20}. Web SSO delegates trust to centralized Identity Providers (IdPs, e.g., Google~\cite{google_signin_2025}), which authenticate users and assert identity to Service Providers via browser redirects.
However, Web SSO requires active IdP sessions; otherwise, users must undergo a multi-step login or registration process, increasing friction.

\noindent \textbf{MNO-based SSO (MSSO)} (also termed one-tap authentication) provides a streamlined login experience~\cite{gsma_china_mobile_identity_2017}. 
\hjs{Despite relying on a similar trust delegation mechanism, MSSO differs from Web SSO in that it shifts the IdP role from an independent web service to the MNO, thereby further extending the trust delegation chain.}
MNOs use mobile data sessions (e.g., 4G/5G) to bind network connectivity to subscriber identity, making the network itself a trust anchor and removing the need for third-party interaction.
Users only need to tap ``verify'' once to complete \hjs{authentication}.
MSSO has therefore been widely adopted, and all three major Chinese MNOs now support it in a mobile market with over 1.8 billion subscriptions~\cite{miit_2025_telecom_statistics}.


Originally deployed in thousands of native mobile apps (e.g., TikTok, Alipay)~\cite{DBLP:conf/dsn/ZhouHCNLG22, gsma_china_mobile_identity_2017, china_mobile_auth_practice_2018}, MSSO has been extended to the open web, with major services such as Taobao.com and NetEase Cloud Music adopting Web-based MSSO login flows.
In this expansion, providers directly reused native-app SDK architectures on the web without fully considering differences in security properties. In native apps, authentication logic is embedded in proprietary SDKs and compiled into binaries, making the execution environment \hjs{relatively opaque.}
However, on the web, the client-side logic runs as JavaScript in the browser. Sensitive data such as credentials, API parameters, and tokens are exposed in readable code, making extraction easier than from compiled binaries.
Prior work~\cite{DBLP:conf/dsn/ZhouHCNLG22} focuses on native apps, but our study targets the web, where the transparent execution environment changes the threat model.
\textit{Web-based MSSO thus lowers the barrier to inspecting and exploiting authentication flows, creating a distinct attack surface.}

\subsection{Preliminary Study of Web-based MSSO}
\label{sec:preliminary_study}

\hjs{Since the deployment architecture and operational workflow of Web-based MSSO\footnote{From this section onward, unless otherwise specified, MSSO refers to Web-based MSSO in the following text.} remain poorly understood}, we first survey mainstream MSSO providers, empirically reconstruct its authentication pipeline, and characterize the MSSO provider landscape, laying the groundwork for the threat model presented in Section~\ref{sec:delegated_risk}.

\noindent \textbf{Seed MSSO Provider Collection.}
\ignore{
To systematically characterize this architectural shift and its security implications, we first mapped the real-world landscape of MSSO service providers and their integration patterns. We surveyed major search engines (e.g., Baidu, Google) and MNO implementation documentation ~\cite{china_mobile_2026_dev, china_telecom_2026_dev, aliyun_2026_pnvs, china_unicom_2026_dev} using keywords (e.g., ``one-tap authentication'') derived from the MSSO mechanism's description to identify its service providers. Through manual analysis, we confirmed 12 providers (see Table~\ref{tab:ecosystem_providers}), which typically offer MSSO capabilities via a service API or an SDK. We found that Chinese MNOs are the most prevalent providers.
This finding aligns with prior work~\cite{DBLP:conf/dsn/ZhouHCNLG22} which identified a similar trend for native mobile applications. In addition, we identified MNOs in other countries (e.g., Vodafone, Orange, Telefónica, AT\&T) that offer GSMA Open Gateway–based authentication services with mechanisms similar to MSSO. 
Through this analysis, we confirmed 12 MSSO service providers and manually analyzed their documentation to extract key API patterns and authentication flows, yielding our initial set of 20 \textit{MSSO Seed Endpoints} (key domains and APIs). 
}
To study this architectural shift, we mapped the MSSO service providers and their integration patterns. 
We searched major engines (e.g., Baidu, Google) and MNO documentation~\cite{china_mobile_2026_dev, china_telecom_2026_dev, aliyun_2026_pnvs, china_unicom_2026_dev} using MSSO-related keywords (e.g., ``one-tap authentication'') derived from the MSSO mechanism's description to identify service providers.
Through manual analysis, we confirmed 14 MSSO providers (see Table~\ref{tab:ecosystem_providers} in Appendix~\ref{apx:msso_provider}), which typically expose services via APIs or SDKs. Chinese MNOs dominate the ecosystem, consistent with prior work on native apps~\cite{DBLP:conf/dsn/ZhouHCNLG22}. 
We also identified international MNOs (e.g., Vodafone, Orange, Telefónica, and AT\&T) offering similar GSMA Open Gateway/CAMARA Number Verification services~\cite{gsma_opg09_2024,telefonica_spain_open_gateway_2024,aduna_us_network_apis_2025}.
From documentation analysis, we extracted 20 \textit{MSSO Seed Endpoints}, also listed in Table~\ref{tab:ecosystem_providers}, covering key HTTP API patterns and authentication flows.

\begin{figure*}[t]
    \centering
    \includegraphics[width=0.85\linewidth]{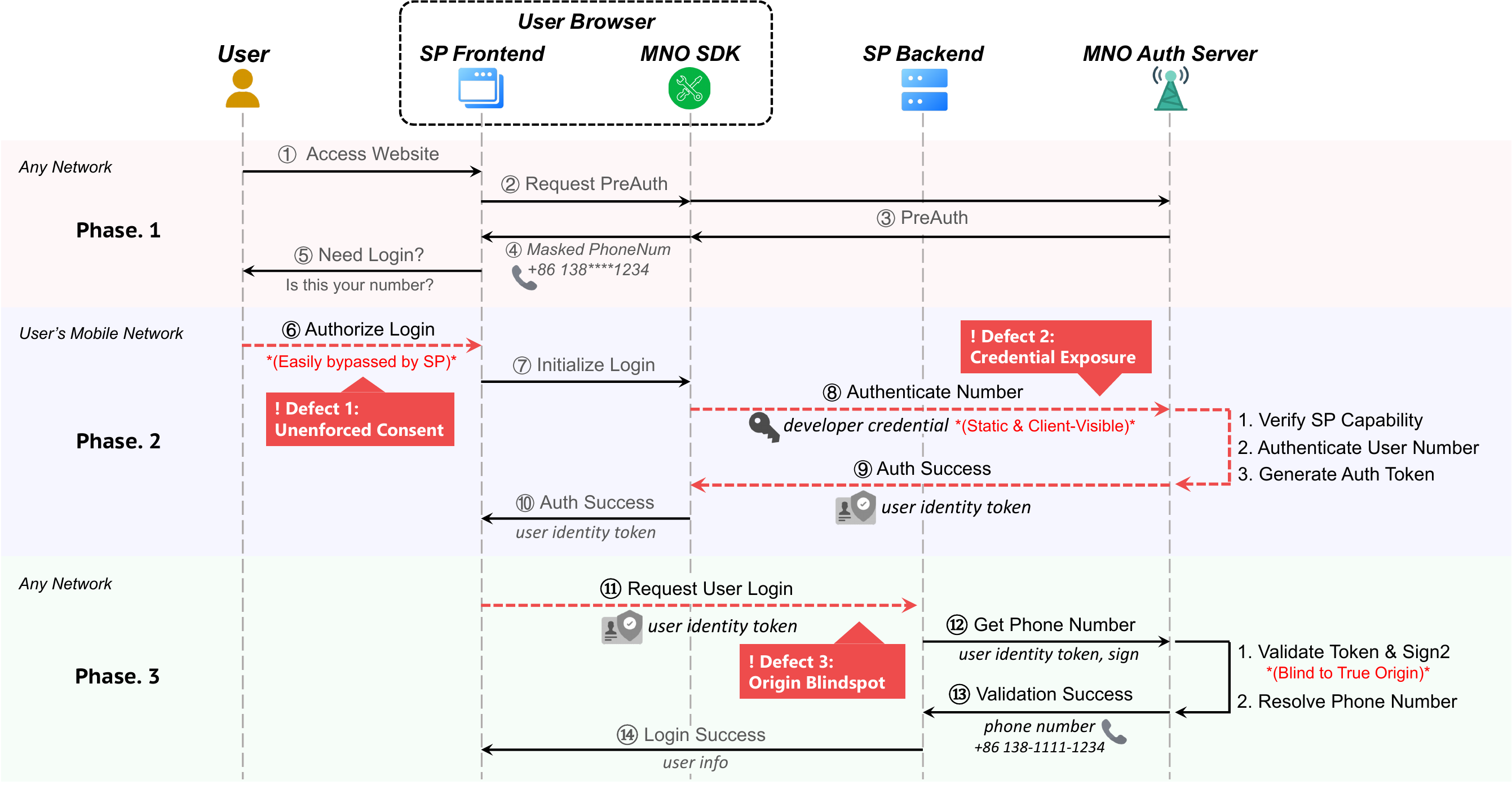}
    \caption{The Three-Phase Workflow of Web-Based MSSO}
    \label{fig:msso-workflow}
\end{figure*}

\noindent\textbf{Authentication Pipeline.}
Based on these seed endpoints, we further identify websites that enable MSSO services, aiming to analyze real-world deployments and uncover the MSSO authentication workflow.
To achieve this, we used these extracted API patterns to perform active pattern matching against the websites on the Secrank Top1M list~\cite{DBLP:conf/uss/XieTZLLD022}, whose Fully Qualified Domain Name (FQDN)-level popularity ranking is ideal for our goal, since MSSO integration status often varies across different service FQDNs within the same Second-Level Domain (SLD). 
We confirmed service enablement by actively loading the webpages for these 1M domains, capturing all network traffic, and identifying whether HTTP requests were issued to our \textit{MSSO Seed Endpoints}.
\label{sec:empirical_study}
Through this process we successfully identified 225 such sites actively utilizing web-based MSSO. For these confirmed sites, we then performed JavaScript call-stack analysis, tracing these requests to their originating scripts. By inspecting the script logic, we derived the parameter handling approach, data-passing mechanisms, and the precise sequence of the real-world authentication flow.
Synthesizing this documentation analysis and empirical validation, we characterized the \textit{delegated trust model} in MSSO and generalized the following three-phase workflow for web-based MSSO authentication, as shown in Figure~\ref{fig:msso-workflow}.
Notably, real-world implementations may vary across different identity providers.
For instance, some MNOs employ a two-stage token process where an initial \textit{accessToken} is required to request the final \textit{User Identity Token}\hjs{; this difference} is further analyzed in Section~\ref{sec:prevalence_of_defect_1}.
Despite minor implementation differences, the core workflow—comprising three phases—remains largely consistent. Our subsequent analysis focuses on these common phases to identify security risks (as discussed in Section~\ref{sec:threat_model}).

\noindent $\bullet$ \textit{Phase 1: Pre-authentication.} 
Operating over the user's mobile data network, SP's frontend fetches a masked version of user's phone number (e.g., 138xxxx1234) from MNO. This step confirms service availability and prepares the user interface, as shown in Figure~\ref{fig:web-msso-demo}. 

\noindent $\bullet$ \textit{Phase 2: Token Issuance.} Following explicit user authorization (e.g., clicking \hjs{the} login button in Figure~\ref{fig:web-msso-demo}), the SP initiates an MSSO authentication request to the MNO over the user's mobile data network. This request carries the static \textit{developer credentials} (obtained from the MNO when purchasing the service) as verification parameters. The MNO first verifies these \textit{developer credentials} to confirm the SP is authorized to initiate authentication. Once the SP's capability is confirmed, the MNO validates the user's identity based on their active data session. The MNO then encodes the verified identity into a short-lived \textit{User Identity Token} and returns it to the SP's frontend over the same mobile data network.

\noindent $\bullet$ \textit{Phase 3: Backend Verification.} In the final phase, the SP's frontend relays the received \textit{User Identity Token} to its own backend. This triggers a server-to-server communication where the SP backend exchanges the \textit{User Identity Token} with the MNO's server to retrieve the user's full phone number. Unlike the first two phases, this server-to-server verification no longer relies on the user's mobile data session, instead completing authentication through a back-end token exchange. The SP backend then establishes a user session and returns the login result to the frontend.

\noindent \textbf{MSSO Service Provider.}
Building on seed data collection, we further analyze the roles of providers offering MSSO services. In principle, only MNOs possess the inherent capability to perform network-level identity verification based on mobile traffic. However, as developer demand for simplified multi-carrier integration grew, third-party providers entered the workflow as intermediaries, offering this service through a unified interface. Because these two types of providers differ fundamentally in their deployment approaches and trust assumptions, we introduce them in two categories (see Table~\ref{tab:ecosystem_providers} in Appendix~\ref{apx:msso_provider} for the full breakdown).

\noindent $\bullet$ \textit{MNOs as Direct Providers.} These are the MNOs (e.g., China Mobile). As the foundational providers, MNOs act as the ultimate identity authority. They exclusively control the network-level mapping between a user's active data session and their verified subscriber identity (i.e., their phone number). They leverage this unique and proprietary capability to offer authentication directly to SPs. 

\noindent $\bullet$ \textit{Third-Party Resellers.} 
Internet service providers (e.g., Alibaba Cloud, Netease) lack authoritative network-level identity verification and are therefore not identity sources of truth.
Instead, they act as service brokers between MNOs and SPs. As each MNO's proprietary service is incompatible and typically only authenticates its own subscribers, developers face severe fragmentation. 
To address this fragmentation, resellers purchase authentication capabilities from multiple MNOs and package them as a unified JS SDK or API for developers, abstracting away the fragmented MSSO provider landscape.
These unified interfaces detect user's mobile network environment and route requests to corresponding MNO endpoint, allowing one integration to serve users across MNOs.

Despite their different deployment approaches, both provider types share the same client-side delegation pattern: the SP's frontend must embed credentials in JavaScript and follow the three-phase authentication flow. 
This shared pattern, as we formalize in Section~\ref{sec:threat_model}, gives rise to a \textit{Trust Hijack Threat}, where delegated trust relationships in MSSO can be hijacked, resulting in unauthorized disclosure of users' verified identities.

\section{Trust Hijack Threat in Web-based MSSO}
\label{sec:threat_model}
Building on the three-phase MSSO workflow characterized in Section~\ref{sec:preliminary_study}, we examine trust delegation and validation across the user, SP, and MNO.
Using the collected network traffic and JavaScript evidence from confirmed MSSO-enabled websites, we reconstruct the trust chain and pinpoint where its security assumptions break in practice.
\hjs{Based on this analysis, we identify three \textit{Trust Defects} in the delegated trust model that collectively give rise to an overarching \textit{Trust Hijack Threat}, in which an adversary hijacks the trust relationships embedded in the MSSO architecture to silently exfiltrate user identity. This threat manifests concretely as the \textit{One-Click-to-Leak (OCL)} attack, a specific exploitation form where a single webpage visit suffices to leak the user's private identity.}

\label{sec:delegated_risk} 
\sssection{Delegated Trust Defect in MSSO}
As detailed in Section~\ref{sec:preliminary_study}, MSSO relies on a \textit{delegated trust model}. 
This model transforms the user into a passive authorizer, thereby sidelining them from the core verification loop. The burden of verification thus shifts entirely to the protocol's intermediaries: the SP and the MNO. This shift creates a dissociation of intent and verification and introduces critical vulnerabilities. We identify three defects specific to this \textit{trust model} from our empirical study, as shown in Figure~\ref{fig:msso-workflow}.

\noindent $\bullet$ \textit{Trust Defect 1: Unenforced User Consent.}
In traditional SSO, user consent is securely anchored and verified on IdP's own domain. In contrast, although MSSO nominally includes a UI-based confirmation step, this mechanism lacks verifiable enforcement by the MNO at the protocol level. By allowing an SP with API privileges to use SP-side JavaScript to bypass the interactive UI and directly invoke the authentication API, this architecture violates the federated-identity requirement that identity-attribute release be governed by an authorized-party decision, typically subscriber consent in public-facing transactions~\cite{nist_sp800_63c_4}. Critically, the MSSO protocol provides no cryptographic or protocol-level binding to prove that a human user actually confirmed the action, meaning the MNO has no reliable way to distinguish a genuine user click from a programmatic API invocation. The MNO blindly trusts the SP's API requests, allowing the authorization flow to be maliciously initiated without the user's actual permission. The same defect exists in native one-tap authentication, where apps may obtain tokens before displaying the SDK's authorization interface~\cite{DBLP:conf/dsn/ZhouHCNLG22}. Therefore, this cross-platform defect is not specific to either platform but is inherent in delegating user-consent enforcement to untrusted client-side code.

\noindent $\bullet$ \textit{Trust Defect 2: Static Credential Exposure.}
To authenticate the SP, the protocol relies on static \textit{developer credentials} (e.g., \texttt{appid} and \texttt{appkey}), which developers embed in client-side assets such as JavaScript and JSON configuration files. As analyzed in Section~\ref{sec:background}, the open architecture of the web offers no effective concealment for these embedded credentials. Our analysis finds developer credentials in readable scripts and browser-visible traffic, allowing an adversary to collect reusable SP-authentication material directly from web assets. The same defect exists in native apps, where fixed \texttt{appid} and \texttt{appkey} values can be recovered from compiled binaries through reverse engineering~\cite{DBLP:conf/dsn/ZhouHCNLG22}. Together, these findings show that client-side distribution exposes reusable developer credentials across platforms, while readable web sources lower the extraction barrier.

\noindent $\bullet$ \textit{Trust Defect 3: Origin Validation Blindspot.}
The MNO lacks contextual awareness to reliably verify the legitimacy of terminal SP and true origin of a request in affected flows. In Reseller-mediated flows, MSSO authorization is mediated by Reseller's \textit{developer credentials} and provider-side routing rather than a verifiable binding between the MNO-facing request and the downstream SP. This creates an origin-validation blindspot: a request carrying valid credentials can appear legitimate even when the actual initiating context differs from that of the registered SP. This blindspot is further compounded by the Third-Party Reseller model, where the MNO observes the trusted Reseller's identity rather than the downstream SP's true origin; we measure the prevalence of this pattern in Section~\ref{sec:prevalence_of_defect_3}. The same class of risk could arise in an app deployment if an intermediary authenticates to the MNO on behalf of downstream apps without preserving a verifiable binding to each app, a deployment pattern not reported in prior native one-tap studies~\cite{DBLP:conf/dsn/ZhouHCNLG22,DBLP:journals/tifs/CuiCFB23}. More generally, Trust Defect 3 shows that authenticating an intermediary does not establish the identity of the terminal client unless the delegation is bound end to end. Without such binding identity assertions can escape their intended app, SP, or origin context.

\sssection{Trust Hijack Threat}
\label{threat_model}
\hjs{Building upon our identification of these Trust Defects, we define the overarching threat enabled by their combination as the \textit{Trust Hijack Threat}.}
Our threat model targets MSSO over mobile data.
This scope is broad in practice, as mobile data networks provide population-scale coverage and routinely support mobile web access~\cite{miit_2025_telecom_statistics,itu_facts_figures_2025}.
In this threat, an adversary exploits the trust defects to hijack the delegated trust relationships and execute a hidden private information theft attack that specifically targets the user's phone number and verified identity.

\sssection{\hjs{One-Click-to-Leak (OCL) Attack}}
\hjs{OCL is a concrete attack of the \textit{Trust Hijack Threat}.}
In an OCL attack, a user's mere visit without any login action can be sufficient to leak their private identity (e.g., phone number) to a malicious site and even de-anonymize their browsing activity across the web (see Section~\ref{sec:exploitation}). 
Specifically, the adversary's strategy is to impersonate a legitimate SP, leveraging the victim's mobile data session to request a \textit{User Identity Token}, and then relay this captured \textit{User Identity Token} to the real SP's legitimate verification endpoint to resolve the victim's private identity.
To achieve this goal, the adversary only needs to operate a standard web entry point (e.g., a malicious site or a compromised third-party script) and meet two prerequisites. First, they must obtain the necessary \textit{developer credentials} (e.g., \texttt{appid} and \texttt{appkey}) of a legitimate SP, a step made possible by \textit{Trust Defect 2} (Static Credential Exposure), which structurally forces these credentials to be statically embedded in transparent web assets. Second, the adversary must be able to exfiltrate the captured \textit{User Identity Token} and use it to resolve the phone number. \textit{Trust Defect 3} (Origin Validation Blindspot) directly enables this, as it permits the \textit{User Identity Token} to be replayed from the adversary's own backend across a different origin and session. Furthermore, \textit{Trust Defect 1} (Unenforced User Consent) compounds the threat by allowing the adversary to trigger the entire authentication flow without the user's actual permission. 
Notably, an OCL attack does not require the adversary to possess any prior knowledge of the victims' identities or phone numbers. The threat model represents a universal, mass-exploitation strategy. By deploying a script that integrates leaked credentials spanning various MNOs, the adversary can dynamically probe and adapt to the specific carrier of any visiting user. Consequently, the adversary can readily scale this silent \hjs{identity-stealing} attack against a broad population, \hjs{potentially} affecting any user who accesses the entry point via a mobile data session.

To execute this attack, the adversary proceeds through the following four steps.

\noindent \textit{(1) Luring the Victim:} The adversary lures unsuspecting users to a standard webpage (e.g., a seemingly normal site) that secretly hosts the malicious script. The attack triggers as long as the victim accesses this page via a mobile data network.

\noindent \textit{(2) Covert Token Request:} The script silently executes, dynamically adapting to the user's MNO, and uses the corresponding stolen \textit{developer credentials} to trigger a background token request.

\noindent \textit{(3) Illicit Token Issuance:} The MNO, validating only the stolen \textit{developer credentials} and unable to verify the request's origin, cannot distinguish the malicious invocation. It consequently issues a valid \textit{User Identity Token} and delivers it to the adversary.

\noindent \textit{\hjs{(4) Attack Completion:}}
Finally, the adversary's backend relays the captured \textit{User Identity Token} to the legitimate verification endpoint (operated by the SP or MNO). This action resolves the \textit{User Identity Token} into the victim's full phone number and finishes the login action, completing \hjs{the} data leakage and even account hijack.

Therefore, a non-powerful adversary who cannot break standard cryptography (e.g., TLS) or compromise the victim's device can execute this attack by exploiting the Trust Defects in MSSO. A single, unsuspecting visit to the adversary's page over a mobile data network can expose the victim's private identity. To confirm the threat's practical feasibility, we conducted controlled experiments using author-controlled SIM cards, mobile-data sessions, and phone numbers. We captured \textit{User Identity Token} values generated through real MSSO flows in these sessions and submitted them from separate sessions to the same SP's legitimate verification endpoint, which returned the corresponding verified phone numbers.

\section{OCL Detection}
\label{sec:methodology}
To detect and measure the impact of trust hijack threats in real-world websites, we design and implement \textit{OCL Detector}. It is a hierarchical framework that leverages passive DNS and search engine data to identify MSSO-enabled websites, and further performs active analysis of their authentication behaviors and potential malicious script threats.

\subsection{OCL Detector: Challenge, Insight, Overview}

\noindent \textbf{Detection Challenge.}
Detecting and measuring the trust hijack threats at a web-wide scale presents substantial challenges. 
First, existing malicious website detection methods target generic web-abuse signals~\cite{DBLP:journals/access/ZieniMC23, DBLP:conf/uss/SoskaC14, DBLP:conf/ladc/McGahaganBPC19}, whereas OCL-relevant behavior appears in MSSO-specific authentication traffic, including silent API invocations and \textit{User Identity Token} exfiltration. Second, brute-force crawling at web scale is computationally intractable. Third, monitoring real user authentication flows at the network layer would be costly and privacy-invasive.

\noindent \textbf{Detection Insights.}
Our preliminary study revealed three consistent MSSO deployment properties that guide our detection strategy.

$\bullet$ \textit{Insight 1: MNO as the Authentication Anchor.} The MNO acts as the definitive authentication anchor: every valid MSSO flow must eventually contact an MNO server, either directly or through a Reseller, to resolve the user's identity. This lets our detection focus on MNO interactions and their correlated entities rather than probing the entire web.

$\bullet$\textit{Insight 2: Stable API Endpoints.} 
MNO API endpoints exhibit stable, recognizable patterns in their domain names, URL structures, and query parameters, serving as reliable signatures for MSSO traffic. We use the endpoint domains as \textit{MSSO Seed Domains}, enabling a DNS-based correlation that drastically narrows our search scope.

$\bullet$\textit{Insight 3: Observable Frontend Interaction.} Because MSSO relies on the user's mobile data session for identification, the process is inherently frontend-driven, making the resulting network requests and scripts observable via targeted dynamic analysis.

\begin{figure}[tb!]
    \centering
    \includegraphics[width=1\linewidth]{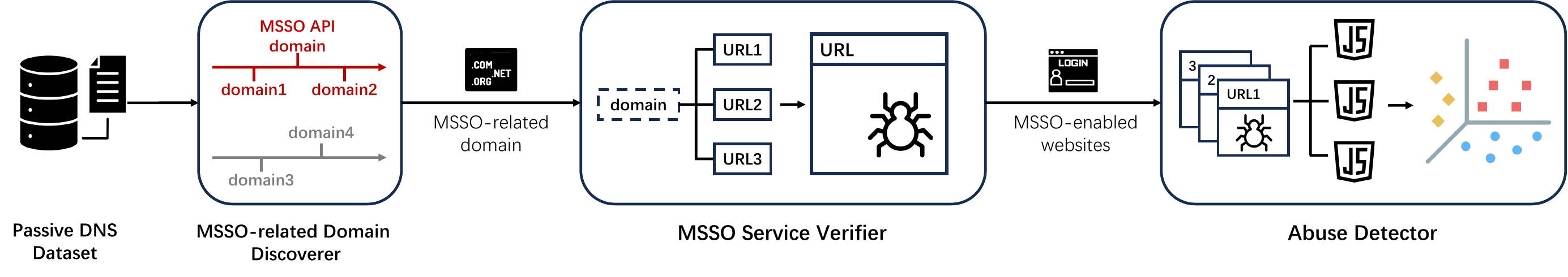}
    \caption{Workflow of \textit{OCL Detector}}
    \label{fig:ocl-detector}
\end{figure}

\noindent \textbf{Overview of OCL Detector.}
Guided by these insights, we designed and implemented \textit{OCL Detector}, a hierarchical detection framework shown in Figure~\ref{fig:ocl-detector}. OCL Detector contains two functional layers. The MSSO identification layer consists of the \textit{MSSO-related Domain Discoverer}, which performs large-scale passive DNS (pDNS) analysis to identify candidate domains correlated with \textit{MSSO Seed Domains}, and the \textit{MSSO Service Verifier}, which translates these domains into candidate URLs and performs targeted dynamic analysis to confirm MSSO integration while collecting evidence. The abuse-analysis layer then uses the collected JavaScript and call-stack evidence to identify suspicious upstream scripts and attribution candidates.

\subsection{MSSO-related Domain Discoverer}
This module identifies \textit{candidate domains}: domains whose DNS resolutions co-occur with known \textit{MSSO Seed Domains} and are therefore likely associated with MSSO-enabled page loads. We use passive DNS (pDNS) because it records query-response pairs observed at recursive DNS resolvers, providing a proxy for real-world browsing through domain resolution patterns~\cite{weimer2005passive,DBLP:conf/ndss/BilgeKKB11,DBLP:conf/uss/AntonakakisPNVALD12}. The analysis proceeds in four steps: constructing per-client query sequences and filtering noisy sources, identifying domains co-occurring with our seed domains, scoring associations with a time-decay model, and aggregating scores across clients with entropy-based weighting.

\sssection{Query Sequence Construction and Filtering}
We construct per-client domain access sequences by grouping DNS queries by source client IP and sorting them by timestamp. Because a single client IP may represent many users behind a large NAT or proxy, its mixed queries can dilute later correlation signals. Following prior pDNS filtering practice~\cite{DBLP:conf/eurosp/LiuLZLDLA0HJZ0Z19}, we remove clients associated with an abnormally large number of distinct FQDNs per day.

\sssection{Domain Co-occurrence Analysis}
On the filtered sequences, we use a 2.0-second sliding window to capture DNS co-occurrences within the same page-load context. This follows our threat model: a candidate domain and its subsequent \textit{MSSO Seed Domain} call should originate from the same page load, so their DNS resolutions should occur within a brief interval consistent with modern webpage load-time targets such as the 2.5-second LCP threshold in Google's Core Web Vitals~\cite{google_core_web_vitals_2025}. For each query $Q_i$, we pair it with subsequent queries $Q_j$ where $timestamp(Q_j)-timestamp(Q_i)\le 2.0\,\mathrm{s}$. Pairs containing both an \textit{MSSO Seed Domain} and a candidate domain are passed to the scoring model.

\sssection{Association Scoring Model}
For each client IP \textit{ip}, we quantify the association between a candidate domain \textit{C} and an \textit{MSSO Seed Domain} \textit{S} via two metrics, adapting a model from prior work~\cite{DBLP:conf/eurosp/LiuLZLDLA0HJZ0Z19}. \textit{Support}, $Supp_{ip}(C) = N_{C,ip}/N_{ip}$, measures \textit{C}'s prevalence in the client's query sequence, where $N_{C,ip}$ and $N_{ip}$ denote the number of queries for \textit{C} and all queries from \textit{ip}, respectively. \textit{Confidence}, $Conf_{ip}(S \Rightarrow C)$, captures how reliably \textit{C} co-occurs with \textit{S}, weighted by temporal proximity. For each co-occurrence in a window \textit{w}, we apply a decay factor $d(S,C,w)=2^{-\lambda|pos_{S}(w)-pos_{C}(w)|}$ that rewards tighter co-occurrences. Summing decay scores across all windows and normalizing by $N_{S,ip}$ yields:
$$ Conf_{ip}(S\Rightarrow C)=\frac{\sum_{w\in W_{ip}}d(S,C,w)}{N_{S,ip}} $$

\sssection{Client-weighted Score Aggregation}
A candidate domain appears in sequences from many clients with varying noise levels. 
We aggregate per-client scores using entropy-based weights: each client's contribution is down-weighted by the Shannon entropy $H(ip)$ of its queried-domain distribution via $w = \frac{1}{1+H(ip)}$, so high-diversity IPs (likely large NATs or proxies) receive lower weight.
The final global score for \textit{C} is the weighted average of all per-client $(Conf_{ip}, Supp_{ip})$ pairs. We additionally discard any candidate resolved by three or fewer client IPs to remove spurious correlations.

The output is a ranked list of candidate domains correlated with \textit{MSSO Seed Domains}, serving as input for the next module.

\subsection{MSSO Service Verifier}
\label{sec:msso_verifier}
This module translates candidate domains into concrete URLs and verifies MSSO integration through dynamic probing.

\noindent\textbf{URL Translation and Normalization.}
We translate candidate domains into URLs using our \textit{website relationship graph}, a large-scale database from a major search engine provider that indexes webpages and models inter-site relationships. Candidate domains fall into two types. For content-providing sites (e.g., \texttt{shopping.example} \texttt{.com}), we retrieve URLs directly from the graph. For resource-providing sites (e.g., \texttt{sdk.example.js.com}) that only supply embedded JavaScript SDKs or \texttt{iframe} content, we use linkage data (\texttt{outerchain}, \texttt{redirect}, \texttt{iframe}) to trace these resources back to the embedding content-providing sites.

The resulting URL list is large and highly redundant; we de-duplicate and sample up to 10 representative URLs per FQDN, prioritizing low path similarity, to reduce probing overhead while preserving coverage. This sampling is applied after FQDN-level candidate expansion and prioritizes path diversity; our evaluation below checks its coverage against the independently confirmed Secrank Top 1M ground-truth set.

\noindent\textbf{MSSO Service Verification.}
We verify MSSO integration by visiting each candidate URL with headless browsers that simulate a mobile network environment and monitoring the resulting traffic. A URL is confirmed when request or response domains and paths match service-signature rules derived from \textit{MSSO Seed Domains} and API patterns (e.g., \texttt{/getPreUrl.do}). When a signature is triggered, we record the timestamp, matched rule, matched request/response URL, and the probed root URL in the \texttt{security alert log}. The module outputs verified MSSO-integrated URLs, the \texttt{security alert log}, and raw evidence for downstream analysis, including HAR files, screenshots, JavaScript resources, and call stack reports. Implementation details of our probing environment are provided in Subsection~\ref{sec:implementation}.

\subsection{Abuse Detector}
\label{sec:abuse_detector}
To understand and trace high-risk activities within the MSSO ecosystem, we design an Abuse Detector.
Directly observing user-data exfiltration in transit is impractical because attackers may encrypt or obfuscate stolen identifiers before transmitting them to backends for server-side resolution. Instead of proving final exfiltration on every site, we target the reusable distribution layer: shared upstream scripts that trigger MSSO calls across downstream sites.
\hjs{
We perform script analysis via clustering. We encode each script using 24 scalar features across three categories:
(1) \textit{General Static Features}, (2) \textit{Provider Features}, and (3) \textit{Specific Behavioral Features}. We further include an averaged path TF--IDF vector and a binary domain multi-hot vector. This representation quantifies the scripts' intrinsic properties, deployment context, and observed runtime actions. The complete feature set is listed in Table~\ref{tab:all-features} in Appendix~\ref{sec:appendix_features}.
}

The final clustering result includes both large, well-defined clusters (e.g., benign libraries, official MNO SDKs) and anomalous outliers, providing the candidate set for downstream abuse analysis.
We treat the resulting clusters as attribution candidates rather than standalone proof of exfiltration; Sections~\ref{sec:abuse_clustering} and~\ref{sec:abuse_cases} validate them through upstream-provider analysis and case studies.

\subsection{Implementation}
\label{sec:implementation}
\sssection{Dataset}
Our study relies on two datasets.
(1) \textit{Passive DNS Logs:} We obtained a large-scale Passive DNS (pDNS) dataset from a public DNS service provider \hjs{in China}, covering its recursive resolver traffic over our study period, at around 500 billion DNS requests and 550 million FQDNs per day.
(2) \textit{Website Relationship Graph:} We use a large-scale graph \hjs{from a search engine provider in China} that integrates webpage snapshots, DNS records, and search engine logs, including domain-to-domain links, site-to-site resource inclusions, and mappings between URLs, sites, and domains.

\sssection{Probing Environment and Evidence Collection}
Our \textit{MSSO Service Verifier} uses a cluster of \texttt{Puppeteer} headless browsers (up to 96 parallel instances via \texttt{Puppeteer-Cluster}), each configured with an Android \texttt{User-Agent}, mobile screen resolution, and mobile network proxy to satisfy MSSO prerequisites. To capture evidence, we use \texttt{Puppeteer-Har} to generate HAR files and implement a \textit{dynamic instrumentation probe} that wraps native browser networking functions (\texttt{window.fetch}, \texttt{XMLHttpRequest}). When an MSSO API rule is triggered, the probe captures the complete JavaScript call stack via a temporary \texttt{Error} object's \texttt{.stack} property.
\hjs{To validate this probing environment, we performed controlled configuration checks on MSSO login pages independently confirmed during our preliminary study and spanning MNO-operated and Reseller deployments. We sequentially switched among different MNO data connections and varied four classes of conditions: network routing, client profile, browser state, and automation settings, repeating each condition three times. The results confirmed that our configured environment covered the conditions required to exercise MSSO authentication flows across both deployment types.}

\sssection{Clustering Algorithm}
\hjs{
We apply z-score scaling to each of the 24 scalar features across scripts. For each script, we average the L2-normalized TF--IDF vectors of its observed paths and use a binary multi-hot vector for its source domains. We concatenate the three blocks without further scaling. We then apply DBSCAN with Euclidean distance, $\epsilon=3.2$, and \texttt{min\_samples}=20. We select $\epsilon$ from the transition in the sorted 20-point-neighborhood k-distance curve. The adjacent settings $\epsilon=3.0/3.2/3.4$ yield 710/586/528 candidates, with noise-set Jaccard similarities of 0.83/1.00/0.90; varying \texttt{min\_samples} from 10 to 30 retains a Jaccard similarity of at least 0.84. DBSCAN requires no pre-specified number of clusters and identifies low-density samples as anomalous candidates for downstream analysis. We then characterize these candidates through feature analysis and manual review, and identify those associated with abusive MSSO behavior through upstream-provider attribution and case studies.
}

\subsection{Evaluation}
We evaluate the \textit{OCL Detector} pipeline from two perspectives corresponding to its key outputs.

\sssection{Accuracy of MSSO-Enabled Websites}
For the websites discovered by the \hjs{MSSO} Service Verifier, we assessed both precision and recall. 
\hjs{
For precision, we used a two-stage evaluation that separates domain-level authentication evidence from URL-level live authentication. Because URLs under the same apex domain commonly share an MSSO integration, the first stage sampled at the apex-domain level to avoid overweighting sites with many detected pages. We randomly sampled 305 of the 729 detected apex domains (41.8\%); the corresponding detections comprised 36,742 URLs. The OCL Detector automatically analyzed the captured traffic for these URLs and produced rule-matched HAR evidence, which we then manually reviewed at the domain level. All 305 sampled domains had reviewed matches containing MSSO authentication evidence. 
Since live authentication depends on the specific page and path, the second stage randomly sampled 367 of these 36,742 URLs (1.0\%) and manually attempted MSSO authentication using author-controlled smartphones over mobile-data connections. Of the 367 URLs, 38 were no longer reachable at validation time. Among the remaining 329 URLs, 319 completed valid MSSO authentication flows (96.9\%); three still loaded an MSSO JSSDK but no longer exposed an active authentication flow, while seven fell back to SMS login under their current deployment configuration.
}
For recall, we compared our results against the ground-truth set of MSSO-enabled sites confirmed in the Secrank Top 1M preliminary study (Section~\ref{sec:preliminary_study}) and found full coverage.
These results indicate high precision and high relative recall with respect to our independently confirmed Secrank Top 1M ground-truth set.

\sssection{Validity of Suspicious Script Clusters}
For each cluster, we checked validity through automatic feature analysis and manual review of representative samples. The automated analysis highlights statistical deviations (e.g., content entropy, domain/path patterns, rule triggers), while manual inspection confirms that these signals reflect coherent behaviors rather than noise or coincidental correlations. Detailed analysis is presented in Section~\ref{sec:abuse}.

\section{Real-World Trust Hijack Assessment}
\label{sec:measurement_assessment}

Having formalized the Trust Hijack threat model (Section~\ref{sec:threat_model}) and built our measurement pipeline (Section~\ref{sec:methodology}), we now present our real-world assessment of how widely these Trust Defects manifest in deployed MSSO services. Using the large-scale scans from our OCL Detector, we quantify both the prevalence of MSSO integrations across the web and the extent to which each Trust Defect is exploitable in practice.




\subsection{The MSSO Deployment Landscape}
\label{sec:eco_prevalence}

\sssection{Overall Deployment}
Using our \textit{OCL Detector}, we processed a pDNS dataset and a large-scale website repository from our search-engine collaborator over a one-year measurement period. Using the \textit{MSSO Service Verifier} (Section~\ref{sec:msso_verifier}), we identified and confirmed \textit{116,852} distinct URLs across \textit{729} apex domains that actively integrate MSSO services for authentication.
To contextualize the real-world defect prevalence, we report statistics at both the URL and domain levels. URL-level counts reflect the total exposure surface across all observed page instances, while domain-level counts control for site-size bias: a single large site may contribute hundreds of URLs, disproportionately inflating URL-level percentages. The domain-level metric better represents the fraction of affected service providers. Where the two diverge significantly, we discuss the underlying cause.

\begin{figure}[b]
    \centering
    \includegraphics[width=0.95\linewidth]{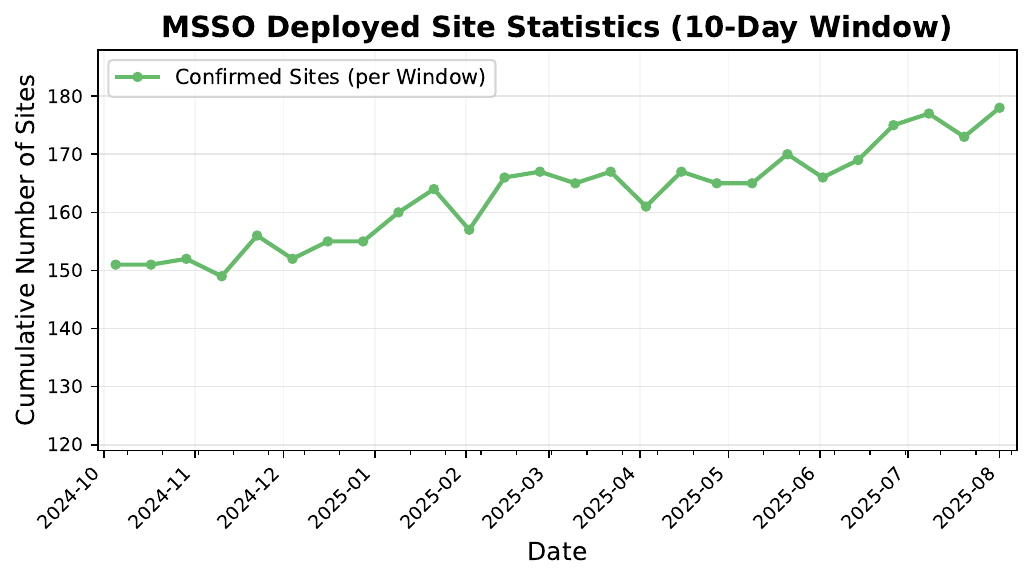}
    \caption{The domain number of unique MSSO-enabled sites observed within discrete 10-day measurement windows.}
    \label{fig:deployment_evolution}
\end{figure}

\sssection{Evolution of Deployment}
Figure~\ref{fig:deployment_evolution} plots the number of unique MSSO-enabled domains observed in discrete 10-day windows. We observe a clear overall upward trajectory, confirming the steady adoption of web-based MSSO. This trend empirically validates our premise that MSSO, once primarily confined to native applications, is actively expanding onto the open web platform. This steady growth underscores the increasing importance of the service and the urgency of addressing its underlying security defects.
The short-term fluctuations are expected artifacts of our pDNS-based methodology. As pDNS data is a passive sample, it only captures sites actively resolved by users within each specific 10-day window. A dip in the count simply means a subset of sites was not present in the observed DNS records for that period, not that they were decommissioned. The overall trend, which aggregates data over time, provides the key insight into the ecosystem's consistent growth.

\sssection{Industry Distribution}
We used the website classification tool provided by our search-engine collaborator (based on title and body semantics) to categorize these websites and analyze the technology’s adoption.
We found MSSO deployed across numerous high-value sectors, demonstrating its deep integration into critical web services. Key sectors include:

\noindent $\bullet$ \textit{Finance and Insurance} (e.g., \texttt{w***bank.com}), where phone numbers are used for high-assurance identity verification.

\noindent $\bullet$ \textit{E-commerce and Travel} (e.g., \texttt{t***bao.com}, \texttt{c***trip.com}), \hjs{where MSSO} reduces friction in checkout and login processes.

\noindent $\bullet$ \textit{MNO Self-provisioned Services} (e.g., \texttt{1***6.cn}), for customer portal access and managing value-added services.

This widespread adoption demonstrates that the potential attack surface is not niche, but spans critical, high-trust sectors of the web.

\subsection{Privileged Tokens Issued Without Consent}
\label{sec:prevalence_of_defect_1}

Against this broad attack surface, we first examine whether the protocol's user consent mechanism holds in practice. \textit{Trust Defect 1 (Unenforced User Consent)}, defined in Section~\ref{sec:delegated_risk}, concerns the MNO's inability to verify that a genuine user action initiated the authentication. As introduced in our workflow analysis (Section~\ref{sec:preliminary_study}), the MSSO process often begins with a \textit{pre-authentication} (or ``pre-fetch'') phase. The benign \textit{goal} of this step is to enhance user experience: it fetches a \textit{low-sensitivity} piece of data, typically a masked phone number (e.g., ``138****1234''), from the MNO. This mask is then displayed in the UI, allowing the user to confirm which identity they are about to log in with \textit{before} they commit to the action.

The core security assumption of this design is that the pre-authentication phase is strictly \textit{low-privilege}. Because it occurs before explicit user consent, it must \textit{only} return non-sensitive data (the mask). The high-privilege token, which can be exchanged for the user's full, verified phone number, must only be generated \textit{after} the user clicks the ``Login'' button, thereby signaling their intent.
However, if this high-privilege token is issued before any explicit user consent, the privilege boundary collapses and the user's verified identifier becomes exposed.

\sssection{Findings}
\hjs{Our analysis of the dynamic network traffic revealed that this core security boundary is frequently violated. We found that 27.1\% (31,654 URLs) of the verified MSSO-integrated pages, corresponding to 24.9\% (182 domains), suffer from a critical \textit{consent-bypass flaw}. In these cases, the MNO's API improperly handles the pre-authentication request. The MNO's server prematurely issues a high-privilege token (e.g., an \textit{access\_token}) during this initial pre-fetch step. This token, which should have been consent-gated, is delivered to the frontend \textit{before} the user has given any explicit authorization (see Figure~\ref{fig:consent-bypass-flaw}).}

\sssection{Implication}
\hjs{
The 27.1\% prevalence shows that pre-consent high-privilege token issuance is already widespread under normal browsing, affecting more than one in four verified MSSO-integrated pages. The fact that these high-privilege tokens are issued before explicit user consent demonstrates that the MNO's token-issuance logic does not reliably depend on a verifiable user action. When combined with the exposure of static \textit{developer credentials} measured next, this behavior becomes directly exploitable: an adversary impersonating an SP can deliberately invoke the same pre-fetch API, completely bypass the protocol's user-consent mechanism, and receive the high-privilege token bound to the victim's identity without requiring the user to click any button.
As demonstrated earlier by our controlled validation in Section~\ref{sec:threat_model}, \textit{User Identity Token} values generated without user interaction in the tested flows were successfully submitted from separate sessions to the same SP's legitimate verification endpoint and resolved to the verified phone numbers associated with author-controlled mobile sessions. 
Together, the large-scale observation establishes the prevalence of pre-consent token issuance, while controlled validation establishes its practical exploitability in the tested flows.
}

\subsection{Leaked and Shared Developer Credentials}
\label{sec:prevalence_of_defect_2}
The consent-bypass above assumes the adversary can already impersonate a legitimate SP. We now examine the feasibility of this prerequisite.
\textit{Trust Defect 2 (Static Credential Exposure)}, defined in Section~\ref{sec:delegated_risk}, examines whether the static \textit{developer credentials} that gate SP authentication can withstand exposure in the web's transparent client-side architecture. In the MSSO protocol, this boundary is enforced by the \textit{developer credentials} (e.g., \texttt{appid}, \texttt{appkey}) held by the SP. The MNO relies on these \textit{developer credentials} to determine \textit{who} is initiating an authentication request.
Our threat model (Section~\ref{sec:delegated_risk}) posits that an adversary must first be able to \textit{impersonate} a legitimate, trusted SP. To do this, the adversary must gain possession of these \textit{developer credentials}. The first question of our practical analysis is therefore: \textit{What is the real-world barrier for an adversary to acquire the SP's \textit{developer credentials}?}

To quantify this risk, we perform this analysis by systematically scanning the \texttt{request} and \texttt{response} bodies within the HAR files collected by the \textit{MSSO Service Verifier}.
We use a set of rule patterns to identify hardcoded MSSO authentication tokens and \textit{developer credentials} (e.g., \texttt{appid} and \texttt{appkey}).
When a match is found, we record a structured entry detailing the website where the leak occurred, the name of the exposed credential, the credential's value, and the MNO we associate with the credential based on its pattern.

\sssection{Findings}
Our analysis reveals that this is a widespread and critical vulnerability.
We found that 69.4\% (81,146 URLs) of the verified MSSO-integrated pages, affecting 62.6\% (457 domains), exposed their \textit{developer credentials}.
In total, we identified 460 distinct exposed \textit{developer credentials}.
The distribution of these leaks spans all major providers: 270 \textit{developer credentials} belonged to China Mobile (CM), 105 to China Telecom (CT), and 85 to China Unicom (CU).
The three MNOs expose developer credentials through structurally distinct channels. CM and CU transmit \texttt{appId} values almost exclusively within JSON response bodies (95.4\% and 100\% of their leaked URLs, respectively), whereas CT exposes \texttt{appId} exclusively through URL query strings in 70.4\% of its leaked URLs. This distinction carries practical significance: URL-embedded credentials are passively recorded in browser history, proxy logs, and \texttt{Referer} headers, making credential recovery possible even without inspecting response bodies, thereby lowering the extraction effort for a potential adversary.
\hjs{This 69.4\% exposure rate is consistent with the structural risk identified in Section~\ref{sec:delegated_risk}: when reusable \textit{developer credentials} are distributed through browser-visible assets or traffic, the web's open architecture makes their exposure a predictable consequence of this integration pattern rather than the result of isolated developer failures.}
\hjs{This widespread exposure can enable reuse of authentication parameters and make malicious requests resemble legitimate SP integrations, weakening the authorization boundary represented by \textit{developer credentials}.}

Furthermore, our analysis of the leaked \textit{developer credentials} reveals a high-risk pattern of credential sharing, where a single set of credentials is used by multiple, often seemingly unrelated, websites.
This practice severely amplifies the impact of a single leak, as it allows one compromised key to unlock services across many websites. We highlight three representative reuse patterns:

\noindent $\bullet$ \textit{Intra-Organizational Reuse:} We found two different \textit{developer credentials} were each hardcoded across 30 distinct websites belonging to the same major e-commerce conglomerate. This pattern indicates that multiple, distinct products from this single entity are all sharing the same credential. It creates a severe amplification of risk: an adversary who obtains this single credential can then impersonate all 30 distinct services, allowing them to leverage one initial compromise across the conglomerate's entire family of products.

\noindent $\bullet$ \textit{Cross-Sector Reuse:} Another single credential was shared by 28 different sites, spanning diverse sectors including health, finance, and travel. This pattern suggests these seemingly unrelated companies are all customers of the same third-party Reseller, a clear real-world manifestation of \textit{Trust Defect 3 (Origin Validation Blindspot)} discussed in Section~\ref{sec:prevalence_of_defect_3}. The immediate risk is a massive blast radius: one leaked credential lets an adversary impersonate 28 companies through their common insecure intermediary.

\noindent $\bullet$ \textit{Widespread Insecure Practices:} We found numerous other instances of sharing, where other \textit{developer credentials} were reused across multiple domains, spanning a wide range of services like insurance and cloud platforms. This demonstrates that such insecure reuse practices are widespread, not isolated to just a few large platforms.

\hjs{Together, the widespread exposure and cross-site reuse of \textit{developer credentials} demonstrate that the SP--MNO authentication boundary is systematically undermined in practice.}

\subsection{Origin Blindspot via the Reseller Ecosystem}
\label{sec:prevalence_of_defect_3}
Even if an MNO attempted to detect impersonation by validating the request's origin, the supply-chain architecture might defeat this defense. We now assess \textit{Trust Defect 3 (Origin Validation Blindspot)}, defined in Section~\ref{sec:delegated_risk}, which arises when the Reseller ecosystem obscures the true initiator of an authentication request. To understand the distribution of the supply chain, we systematically analyzed the major providers in our dataset. Table~\ref{tab:ecosystem_providers} summarizes these key players, categorizing them by their architectural role (MNO or Reseller) and the nature of their service.

\hjs{As shown in Table~\ref{tab:ecosystem_providers}, Reseller model is a prevalent part of the ecosystem. However, this model introduces a distinct origin-validation gap that compounds Defect 3 (the Origin Validation Blindspot). By acting as a trusted intermediary, the Reseller obscures the true identity of the downstream SP initiating the authentication, thereby limiting the MNO's ability to validate the request's true origin.}

In the basic MSSO framework, trust is delegated from the user to the SP. This delegation already creates a gap where the MNO cannot be certain of the user's knowing participation. An attacker who steals an SP's \textit{developer credentials} could thus impersonate that SP, request authentication for a victim, and steal their identity.
\hjs{Theoretically, a vigilant MNO might attempt to mitigate this \textit{protocol-level} weakness by enforcing strict source validation, such as checking the request's origin against the SP's registered domain. However, the introduction of the Reseller ecosystem limits this defense in the analyzed flows.}
As the MNO serves as the ultimate Authentication Anchor, even when a downstream SP (a customer) subscribes to a reseller's service, the authentication process must ultimately be resolved by an MNO-controlled endpoint. The Reseller's SDK mediates this flow, but it culminates in a request to the MNO using the Reseller's identity and \textit{developer credentials}, not the customer SP's.
\hjs{This creates the critical indirection. The MNO's endpoint receives a request that appears valid, as it originates from a trusted partner (the Reseller). However, from the MNO-facing request, the MNO cannot identify the downstream SP. It cannot distinguish a legitimate authentication request from \texttt{shopping.example.com} from a malicious one initiated by \texttt{attacker-site.com}, as both are masked behind the Reseller's single, valid identity. This dynamic creates the Origin Validation Blindspot defined in Section~\ref{sec:threat_model}.}

\hjs{To assess the widespread occurrence of this architectural risk, we analyzed the distribution of provider roles based on the \texttt{Prevalence (Sites)} column in Table~\ref{tab:ecosystem_providers}. Our analysis reveals that a significant portion of the ecosystem uses Reseller-mediated integration: \textbf{\textit{31.8\%}} (232) of the \textit{729} apex domains, corresponding to 19.6\% (22,859 URLs), integrate MSSO services via a Third-Party Reseller.}
\hjs{
This finding demonstrates that the Reseller deployment pattern underlying Defect 3 is not theoretical or niche, but widespread in the wild. In the analyzed flows, request indirection leaves the MNO unable to identify the downstream SP from the MNO-facing request, which can increase the feasibility of the \textit{OCL Threat}.
}

\hjs{The Reseller model is associated with a higher observed credential-exposure rate. Reseller-connected URLs exhibit a credential leak rate of 80.1\%, compared to 66.6\% for sites connecting directly to MNOs (74.2\% vs.\ 57.2\% at the domain level). Reseller SDKs support multiple MNOs, and a single page load can expose credentials from multiple MNOs at once. We observed this multi-MNO co-occurrence on 122 domains, with CM and CT credentials appearing together most frequently (115 domains). The same major e-commerce conglomerate whose intra-organizational credential reuse is documented in Section~\ref{sec:prevalence_of_defect_2} exposes credentials spanning all three MNOs across its 30 Reseller-integrated domains, providing a concrete example of multi-MNO co-exposure. Taken together, these measurements show that Reseller-connected deployments have a higher observed exposure rate and place the Reseller rather than the downstream SP at the MNO-facing authentication boundary.}


\section{One-Click-to-Leak Attack Evaluation}
\label{sec:abuse}

\label{sec:exploitation}

To measure in-the-wild exploitation of the OCL attack, we analyze the attacker supply chain, covering both downstream suspicious scripts and the upstream providers that distribute them. 
We further reconstruct the abuse and monetization pipeline using sanitized backend code and data, obtained through our industry partner and law enforcement, from one representative seized platform.



\subsection{Malicious Provider Discovery}
\label{sec:abuse_clustering}
As discussed in Section~\ref{sec:abuse_detector}, directly observing exfiltration in transit is impractical. We therefore adopt the indirect discovery strategy enabled by our Abuse Detector, combining downstream behavioral analysis to identify suspicious scripts with upstream attribution to trace them to their source providers, revealing the platform-based nature of abusive MSSO exploitation.
%
We first filtered 9,782 unique JavaScript files (from our dynamic probe) for high-risk targets via the Abuse Detector (Section~\ref{sec:abuse_detector}), with feature engineering (Table~\ref{tab:all-features}) and DBSCAN clustering. This clustering produced 586 anomalous outliers, a baseline Cluster~0 containing 8,729 scripts, and \hjs{9} other clusters (benign and suspicious) totaling 467 scripts.

\sssection{Benign Clusters}
Our analysis first successfully grouped the majority of scripts into large, benign clusters with clear, interpretable characteristics. Examples include:

\noindent $\bullet$ \textit{Cluster 3 (Official MNO Scripts):} 
This cluster is characterized by first-party resource loading, high authentication trigger counts, and domains operated by MNOs, identifying these scripts as official operator-provided SDKs and a clear signature of legitimate, high-intent authentication.

    
\noindent $\bullet$ \textit{Cluster 8 (Ad Platforms):} 
This cluster comprises widely distributed third-party scripts, often with versioned URLs, which manual inspection identifies as advertising and analytics platforms; their near-zero authentication trigger rate shows our features effectively distinguish co-located scripts from those actively involved in authentication.

\sssection{Suspicious Usage Patterns}
Analyzing the smaller suspicious clusters and 586 outliers, we identify several distinct high-risk MSSO integration patterns. In particular, by examining the feature dimensions along which outliers deviate, we group scripts with similar risk profiles, yielding the following categories:

\noindent $\bullet$ \textit{Pattern 1: Obfuscated First-Party Logic.} This pattern is defined by scripts loaded as first-party assets, having low prevalence, and exhibiting high content entropy. This profile matches site-specific, custom-developed code that has been intentionally obfuscated. The combination of high relevance to authentication behaviors with obfuscation suggests a deliberate effort by these sites to hide their custom MSSO integration logic, possibly to conceal non-standard data collection or flawed implementations.
    
\noindent $\bullet$ \textit{Pattern 2: Hyper-Active First-Party Integration.} This pattern is also marked by first-party scripts, but its defining characteristic is an extremely high frequency of triggering authentication events. This indicates that MSSO is not just a login option but is a core, high-frequency business function (e.g., triggered on many pages, not just at login). While not inherently malicious, this ``hyper-active'' pattern creates a significantly larger attack surface and increases risk, as the high-privilege authentication process is invoked far more often than necessary.
    
\noindent $\bullet$ \textit{Pattern 3: Covert Exploit \hjs{Platform Candidates}.} This pattern consists of scripts loaded purely as third-party resources and exhibiting low prevalence.
\hjs{They are not official MNO SDKs and display key high-risk traits: high authentication trigger rate (MSSO as sole purpose) and high content entropy (obfuscation). Together, these traits are consistent with third-party Exploit-as-a-Service platforms that package OCL attacks as unofficial covert solutions for downstream customers.}

Finally, this clustering analysis narrowed our investigation from 9,782 scripts to \textbf{\hjs{586 potentially high-risk scripts}}, from the suspicious clusters and anomalous outliers.
\hjs{
Furthermore, we screened all 586 retained resources.
Of the 582 parseable JavaScript resources (the remaining four were JSON-like payloads), we found that 255 (43.8\%) contained at least one obfuscation pattern recognized by the pinned AST-based \texttt{obfuscation-detector} v3.0.0~\cite{baryo_obfuscation_detector_2026}. A conservative structural screen further found non-literal \texttt{eval} or \texttt{Function} construction in 214 resources, explicit debugging-interference structures in 6, and environment-dependent control of a delivery or execution sink in 2.
These measurements characterize observable transformations and code structures within the high-risk set.
}

\sssection{Upstream Provider Attribution}
\label{sec:abuse_attribution}
Having identified 586 potentially high-risk scripts deployed across 564 domains through downstream behavioral clustering, we further trace these scripts upstream to their source providers.
%
We conducted an Upstream Provider Attribution analysis, examining all 729 MSSO-enabled sites identified by our MSSO Service Verifier (Section~\ref{sec:msso_verifier}). For each site, we extract the source providers of every external subresource and rank these providers by the number of MSSO-enabled sites that reference them, thereby pinpointing the entities most strongly correlated with MSSO deployments.
Among the 586 high-risk scripts, 162 (\hjs{27.6\%}) were served by 5 upstream platforms, covering 101 websites.
\hjs{The scripts that were the most anomalous were particularly concentrated among just a few of these platforms. This finding strongly indicates that the observed high-risk MSSO behavior is not the result of sporadic, individual developer error, but rather a platform-based activity distributed through a few upstream providers.}

\subsection{Case Study: In-the-Wild Exploitation}
\label{sec:abuse_cases}

\hjs{We now present evidence of how the OCL attack is exploited in practice. We first describe two representative case studies identified through clustering and attribution analysis, which provide strong script- and provider-level association evidence. We then analyze sanitized backend code and data from one representative seized platform; this backend evidence directly confirms one operational OCL exploitation pipeline and reveals its industrialized scale.}

\sssection{\hjs{Strongly Associated Abuse Patterns}}
\hjs{Cross-referencing the downstream behavioral analysis with the upstream source attribution shows that the suspicious patterns are concentrated rather than isolated. Because final user-data exfiltration cannot be reliably observed in transit, we classify these 101 websites as strongly associated with abusive MSSO exploitation through scripts supplied by 5 upstream platforms. Two representative cases illustrate distinct exploitation strategies.}

\noindent $\bullet$ \textit{Case Study 1: Co-opted Analytics Platform.}
This abuse pattern links to Platform A, a high-frequency upstream provider from our attribution. Its clustered anomalous outlier scripts masquerade as user analytics tools but instead secretly trigger the MSSO pre-fetch API (exploiting the Trust Defects characterized in Section~\ref{sec:measurement_assessment}) on page load, capture the User Identity Token, and exfiltrate it to Platform A's servers, \hjs{enabling user de-anonymization} without consent.

\noindent $\bullet$ \textit{Case Study 2: Malicious Service Chaining.}
This pattern appears in Platform B’s scripts (Cluster 2: Obfuscated First-Party Logic), which chain two Trust Defects: using stolen downstream client \textit{developer credentials} (Trust Defect 2, Section~\ref{sec:prevalence_of_defect_2}) for requests, and calling the vulnerable pre-fetch API (Trust Defect 1, Section~\ref{sec:prevalence_of_defect_1}) to obtain high-privilege User Identity Tokens without user interaction. These tokens are instantly exfiltrated to a third-party data-tracking platform, completing the malicious service chain.






\sssection{The OCL Black-Hat Economy}
\label{sec:blackhat-economy}
\hjs{Our case studies establish strong associations between abusive MSSO behavior and identifiable upstream platforms.} To understand the resulting economy, we reconstruct one representative seized platform's operational and financial pipeline from exploit delivery to monetization.
Specifically, through our collaborator, a leading search engine company, we traced suspicious scripts to upstream platforms using our OCL Detector pipeline (Section~\ref{sec:abuse_clustering}) and reported the findings to law enforcement. The subsequent investigation and takedown operation resulted in the seizure of several such platforms. Our analysis uses sanitized backend code and data from one representative seized platform, with all personally identifiable information removed prior to analysis.





This seized platform operates as a multi-actor criminal enterprise, with roles mirroring legitimate business structures. To clarify these roles within the black-hat economy, we define the following: 1) Victim is the end-user browsing the web and using search engines over a mobile data network; 2) Platform is the black-hat platform providing the exploit as a service; 3) Customer is the malicious website operator who pays the \textit{Platform} to deploy the exploit on their site; 4) Affiliate is the customer who is also incentivized to resell the Platform's user de-anonymization service to other Customers.

\ignore{
$\bullet$ \textit{Victim}: The end-user browsing the web and using search engines over a mobile data network.

$\bullet$ \textit{Platform}: The black-hat platform providing the exploit as a service.

$\bullet$ \textit{Customer}: A malicious website operator who pays the \textit{Platform} to deploy the exploit on their site.

$\bullet$ \textit{Affiliate}: A Customer who is also incentivized to resell the Platform's user de-anonymization service to other Customers.
}

Our analysis reveals the Platform’s three-stage industrial architecture. The \textit{upstream} is a sophisticated Software-as-a-Service (SaaS) platform offering OCL exploits as a product, distributed via a multi-level Affiliate network with industrialized rules (e.g., 1.5 CNY minimum per stolen number, minimum purchase quotas). The \textit{midstream} is an automated two-stage data-fusion pipeline: first capturing Victims’ basic identifiers (phone number, IP, referrer), then enriching profiles by leveraging Customers’ webmaster/analytics credentials to pull and associate Victims’ search keywords. The \textit{downstream} handles monetization via anonymous USDT payments and a fully automated commission system incentivizing Affiliates.

\sssection{Upstream: The SaaS Exploit-Delivery Platform}
%
The Platform’s upstream layer functions as a centralized SaaS provider that productizes exploit capabilities, with an admin panel defining business rules for Affiliates who resell de-anonymization services. Backend scripts enforce this hierarchy by validating credit quotas and price controls when provisioning Customers. When Victims visit Customer sites, malicious scripts identify MNOs via IP geolocation and deploy tailored exploits to extract phone numbers or user identity tokens, while remaining highly resistant to detection:

\noindent $\bullet$ \textit{Cloaking and Authentication:} The script is disguised as a common CDN library. The delivery URL uses unique subdomains and path components (e.g., \texttt{dz37cb.malsite.com/zfww/jqurey.js}). Backend logic confirms that these URL components function as a method to authenticate paid Customers before serving malicious payload. 

\noindent $\bullet$ \textit{Anti-Analysis:} The scripts are replete with anti-analysis checks, such as checking that the request includes a mobile \texttt{User-Agent} and has a valid \texttt{HTTP\_REFERER}. Furthermore, we observed it employs techniques to actively thwart debugging, such as disabling all \texttt{console} functions on the Victim's webpage.

\noindent $\bullet$ \textit{Stealthy Exfiltration:} The client-side payload, dynamically generated by the server, uses a stealthy exfiltration chain. It dynamically creates a hidden \texttt{img} tag and sets the MNO's authentication URL as its \texttt{src} attribute. The image \texttt{onload} event is used as a callback to trigger an \texttt{XMLHttpRequest}, which stealthily sends the captured token or phone number back to the Platform's backend.

\sssection{Midstream: Automated Victim Profile Enrichment}
The Platform's primary value is not just collecting phone numbers, but enriching them, with a two-stage automated pipeline.

\noindent $\bullet$ \textit{Stage 1: Initial Capture.} The MNO-specific exploit scripts perform the first level of enrichment. Upon successfully stealing the phone number, they execute a \texttt{SQL INSERT} query to write the Victim's profile into the main \texttt{data} table. 
It binds the phone number to the Victim's \texttt{referrer} (the webpage they came from), \texttt{IP}, IP-based geolocation, User-Agent, and the Platform's Customer ID (\texttt{username}).

\noindent $\bullet$ \textit{Stage 2: Server-Side Keyword Fusion.} 
%
The Platform enriches this data with Victims’ search keywords via an automated backend task. It queries the \textbf{user} table for Customers who provided webmaster analytics credentials (e.g., usernames, passwords, access tokens), then uses these to periodically fetch visitor logs (IP and search keywords) from the search engine’s analytics service. The backend auto-fuses this data, matching keywords to Victims’ profiles via IP to link their phone numbers with high-intent search queries.

\sssection{Downstream: Automated Monetization and Commission}
With enriched victim profiles in hand, the Platform turns to monetization. To evade financial regulation, it uses cryptocurrency: its Customer-facing payment page directs Affiliates to pay via USDT (Tether) on the TRC20 network, with a static wallet address. Payment is fully automated—a backend script leverages a crypto exchange API to monitor deposits and auto-credit Customer accounts.
Beyond payment processing, the Platform incentivizes distribution through an automated, multi-level-marketing (MLM) commission structure. Our analysis of the backend code shows that when a Customer makes a payment, the script traverses up the Affiliate chain. It retrieves the pre-set cost basis for both the Affiliate and their parent Affiliate, calculates the profit margin on the sale, and immediately credits the parent Affiliate’s account with their commission, converted back into service credits. This automated, pyramid-style commission structure creates a powerful financial incentive for Affiliates to perpetuate and resell the OCL exploit.

The Platform's database shows that it covertly obtained and profiled \textit{14,100 unique Victim phone numbers in just three days}. Based on the Platform's minimum price floor of 1.5 CNY per number, this three-day collection period represents potential illicit revenue of over 21,000 CNY (approx. \$2,900 USD). This confirms that OCL attacks are active, scalable, and highly monetized in the wild.

\section{Discussion}
\label{sec:discussion}
\noindent \textbf{Recommendation.}
To enhance the security of web-based MSSO and protect user privacy, we propose the following mitigations:

\noindent $\bullet$ \textit{Gate Token Issuance with User Verification.} 
MNOs should gate \textit{User Identity Token} issuance on a low-friction user-verification step, rather than relying only on a consent click after displaying a masked phone number. The key requirement is that the user provide information not exposed by the conventional one-tap flow, preventing an automated script from completing authentication alone. One implementation is \textit{Masked-Number Completion Verification}: as illustrated in Figure~\ref{fig:masked-number-completion}, the interface leaves the middle digits in the masked number blank (e.g., \texttt{138xxxx1234}) and requires the user to enter those digits before token issuance. This mechanism adds explicit user participation while preserving MSSO's low-friction experience; MNOs should pair it with server-side rate limiting or temporary blocking to prevent brute-force attempts.

\noindent $\bullet$ \textit{Enforce Strict Privilege Separation.} 
To mitigate the consent-bypass vulnerability (27.1\% of verified MSSO-integrated URLs), MNOs should strictly prevent pre-authentication endpoints from issuing high-privilege tokens and ensure token generation occurs only after explicit user consent via a dedicated authentication endpoint.


\noindent $\bullet$ \textit{Deprecate Static Developer Credentials.} The widespread leakage of \textit{developer credentials} (in 69.4\% of verified MSSO-integrated URLs) demonstrates their insecurity. MNOs are encouraged to deprecate static \texttt{appid}/\texttt{appkey} mechanisms in favor of a more secure model, such as short-lived dynamic tokens fetched by the SP backend.



\noindent \textbf{Responsible Threat Disclosure.}
\hjs{
We provide several repair recommendations and have responsibly disclosed our findings to affected MNOs, SPs, and relevant stakeholders, who have acknowledged this threat and are positively working on remediation.
Encouragingly, following our disclosure, we observed that at least 148 affected sites among the domains identified by OCL Detector had deployed recommended defenses, including safer MSSO implementations such as masked-number completion. This observation demonstrates a practical way to reduce OCL exposure without abandoning MSSO's low-friction login experience.
}

\noindent \textbf{Limitation.}
Despite significant efforts in understanding web-based MSSO, our work still has limitations.
\hjs{First, the datasets we used (PDNS and search engine index data) are sourced from Chinese vendors, introducing geographical limitations to our evaluation results. However, both datasets serve a large user base: the PDNS dataset includes around 500 billion unique DNS requests for about 550 million FQDNs daily, and the search engine data covers over 20 million users with hundreds of millions of indexed pages. Thus, although deployment prevalence and provider practices may differ across regions, our large-scale datasets still reveal the deployment status and real-world risks of a mature web-based MSSO ecosystem, and these findings can generally inform similar MNO-based authentication deployments globally.}
\hjs{
Second, our method for identifying MSSO-enabled websites may have limitations. 
On one hand, data noise may cause bias when correlating PDNS with candidate MSSO-related domains. But, we verified this step using actively collected MSSO Seed Endpoints (Section~\ref{sec:preliminary_study}), with no false positives detected. Websites may also condition MSSO delivery on the client or network environment, which can cause false negatives during dynamic probing. To reduce this risk, we conducted the controlled validation described above and configured the probing environment to preserve the conditions required for ordinary MSSO authentication flows while avoiding unnecessary changes to the browser environment. 
Nevertheless, adaptive cloaking can withhold a payload before collection, while obfuscation and environment-dependent branches can restrict source and runtime analysis to delivered and executed paths. Our traffic signatures and call-stack capture reduce reliance on readable source code, but cannot recover withheld or unexecuted behavior. 
The detected deployment and abuse counts are therefore conservative lower bounds under such evasion, while the positive responses to our disclosures confirm the risk’s widespread impact.
On the other hand, the clustering-based abuse detector leveraging static and dynamic page features may produce biased results. However, we manually validated randomly sampled results from each cluster, finding no ambiguity. Together, the controlled validation and manual review support the observed positive results, while residual evasion primarily risks false negatives.
}

\section{Related Works}
\label{sec:related}
\sssection{Mobile Authentication Security}
\hjs{Prior research has characterized security limitations across the authentication mechanisms that precede MSSO. Studies of passwords assess the security and usability limitations of user-managed secrets~\cite{DBLP:journals/cacm/BonneauHOS15,DBLP:conf/sp/BonneauHOS12}, while studies of MNO-assisted authentication identify SMS OTP interception on mobile devices and weaknesses in MNO authentication procedures for SIM swaps~\cite{DBLP:conf/ndss/LeiNFB21,DBLP:conf/soups/LeeKMN20}. As authentication shifts toward Web SSO, research characterizes IdP--SP architectures and security properties, including formal analyses of OAuth/OIDC protocols~\cite{DBLP:journals/csur/AlacaO20,DBLP:conf/eurosp/MainkaMSW17,FettKS16}. Empirical and automated studies examine security flaws in deployed SP integrations and provider libraries~\cite{WangCW12,DBLP:conf/ccs/SunB12,ZhouE14,DBLP:conf/ccs/Rahat0T22}, while other work investigates account and session compromise and automatic privacy leakage during visits to SPs~\cite{GhasemisharifRC18,DBLP:conf/acsac/WestersMJ24}. These studies focus on Web IdP accounts, protocol tokens, SP integrations, and web sessions, whereas web-based MSSO shifts the identity authority to the MNO, which maps an active mobile-data session to a verified phone identity.
Recent MSSO work identifies vulnerabilities in cellular-network-based one-tap authentication, including unverifiable network authenticity~\cite{DBLP:journals/tifs/CuiCFB23} and the MNO server's inability to distinguish which app initiated authentication, leading to the \textit{SIMulation attack}~\cite{DBLP:conf/dsn/ZhouHCNLG22}. In contrast, we study web-based MSSO at scale and show how leaked developer credentials, consent bypass, and weak origin validation enable OCL.}


\sssection{Phone Numbers as an Abuse Primitive}
Prior work has shown that phone numbers are high-value identifiers for fraud, profiling, and targeted abuse. Studies of telephony fraud and mobile telephony threats show how attackers monetize the reachability and trust associated with phone numbers~\cite{DBLP:conf/eurosp/SahinFGA17,DBLP:conf/ccs/BalduzziGGGA16}, while work on cross-application features and contact discovery shows how phone numbers enable targeted attacks and large-scale user enumeration~\cite{DBLP:conf/ccs/GuptaGAK16,DBLP:conf/ndss/HagenWSD021}. Web phishing represents another major path for identity compromise, relying on social engineering and deceptive interfaces to harvest credentials~\cite{DBLP:conf/uss/HoCGSPSV019,DBLP:conf/uss/LinLDNCLSZD21}. OCL occupies a different point in this attack space: an adversary can operate an ordinary web entry point, reuse exposed developer credentials, and turn a standard mobile-data page visit into leakage of a verified mobile identity. This low-barrier, high-value property makes OCL attractive for monetized abuse, because leaked phone numbers can be immediately linked to web context such as referrers, IPs, and search intent.

\section{Conclusion}
\label{sec:conclusion}
MSSO is an emerging password-free authentication framework leveraging users' active mobile data sessions. Unlike traditional SSO, it shifts the IdP role from web services to MNOs and delegates the authentication anchor to SPs.
\hjs{Through empirical analysis of MSSO mechanisms, we identify three Trust Defects that collectively give rise to a Trust Hijack Threat; OCL is the concrete attack form that stealthily steals login-sensitive data. Collaborating with a leading search engine company, we designed a three-stage detection framework for the first large-scale assessment of the MSSO ecosystem, spanning 116,852 URLs across 729 apex domains and quantifying its prevalence, key providers, and observed Trust Defect manifestations (Trust Defect 1: 27.1\% of URLs exhibit pre-consent high-privilege token issuance; Trust Defect 2: 69.4\% of URLs expose developer credentials; Trust Defect 3: 31.8\% of apex domains use Reseller-mediated integration, the deployment pattern underlying this defect).
We present the first real-world evidence of OCL attacks: our analysis identifies 101 websites strongly associated with OCL attack behavior through scripts served by 5 upstream platforms. Sanitized backend evidence from one representative seized platform also reveals an operational OCL exploitation pipeline that collected 14,100 unique phone numbers within three days and monetized the resulting data. We propose mitigations and have conducted responsible disclosure; together, these efforts help improve mobile authentication security.}

\bibliographystyle{ACM-Reference-Format}
\bibliography{main}

\appendix

\section{Open Science}
\label{sec:open_science}

\hjs{
The code artifact is anonymously available at
\url{https://doi.org/10.5281/zenodo.22733454}.
It contains the source code of \textit{OCL Detector}, including passive DNS correlation, dynamic service verification, script fingerprinting with DBSCAN clustering, and the shared MSSO seed-domain rules. The passive DNS and search-engine data cannot be redistributed under the applicable data-sharing agreements.

We also release a sanitized notebook, aggregate DBSCAN outputs, examples of MSSO flows that adopted masked-number completion following our disclosure, and a controlled end-to-end OCL demonstration using only author-controlled SIM cards, mobile-data sessions, and phone numbers.

For legal, ethical, and safety reasons, we do not release raw measurement datasets or site- and script-level records. We also do not release the HAR archives because they contain live credentials, authentication endpoints, and real interaction flows that could be misused. Instead, we provide anonymized aggregate DBSCAN results and an accompanying sanitized notebook. We do not release raw script fingerprints, JavaScript resources, feature rows, cluster-membership information, affected-site identities, or attack code, as these materials could expose vulnerable MSSO deployments or facilitate attacks. Under our confidentiality agreement, data obtained from the seized platform is used only for anonymized internal analysis.

Therefore, the anonymous repository allows reviewers to examine how each pipeline stage and module interface of \textit{OCL Detector} processes its inputs and produces downstream outputs. 
Although the restricted datasets prevent full reproduction of our measurements, the paper reports all quantitative findings.
The public artifact neither provides live vulnerable deployments nor provides materials that could enable OCL attacks against them. Nevertheless, the sanitized DBSCAN analyses, examples of safer MSSO flows, and controlled attack demonstration provide evidence of the risk and its real-world impact.
}

\section{Ethical Considerations}
\label{sec:ethics}
\ignore{
数据处理：
PDNS中的client IP均是匿名化的；
收录数据基本上也都是一些公开的页面，不存在数据上有ethical风险；
主动请求：
控制请求速率；
在发起请求的服务器上做配置，说明实验探测的目的和联系方式，允许被探测的网站随时退出我们的测量；
滥用案例后台数据的分析：
利用我们的abuse detector能够从百度索引的数据中获得具有风险甚至恶意性的页面，我们对所有监测结果均进行了负责任的披露，即向存在风险的运营商进行披露，目前联通已经积极反馈，其次通过国家CERT机构向这些受影响的页面进行披露；
最后，根据我们的监测结果，我们通过合作的搜索引擎与国家执法部门合作，协作打击治理了一些滥用页面，这些被治理的网站的数据（如后台）均是由我们的合作者进行了匿名化后提供给我们（比如手机号都是加盐哈希值），我们并不直接接触敏感数据。
}


\hjs{
We designed our measurement and validation procedures to minimize potential risks in accordance with the Belmont Report and the Menlo Report~\cite{united1978belmont,kenneally2012menlo}. Our safeguards cover experimental oversight and validation, active probing, data handling and release, and responsible disclosure.
First, all experiments and data handling were conducted under the legal and compliance oversight of our industry collaborator. The probing environment used author-controlled SIM cards and mobile-data sessions. Token-to-phone validation was limited to our own SIM cards, sessions, and phone numbers: we neither redeemed User Identity Tokens belonging to third-party users nor attempted to obtain their phone numbers.
Second, our active probing was limited to publicly indexed pages. The \textit{MSSO Service Verifier} loaded these pages and collected authentication traffic, with requests to each target rate-limited to no more than one per second. The probing servers hosted a notice describing the study, provided a contact email address, and allowed sites to opt out of the measurement at any time. To date, we have received no complaints.
Third, we minimized sensitive-data exposure throughout collection, analysis, and release. The pDNS and search-engine datasets were handled under partner agreements. Our collaborators anonymized pDNS client IP addresses before sharing the data, and we accessed only hashed identifiers. Backend evidence from the seized platform was provided only after sanitization or aggregation, including salted hashing of phone numbers, and we did not access raw sensitive records. We do not release raw HAR logs, lists of affected sites, developer credentials, or backend records because doing so could expose affected services or enable secondary abuse.
Finally, we disclosed identified risks through our industry collaborator to affected MSSO Providers and platforms, and reported affected pages to the national CERT. Our collaborator also reported suspected abuse to national law enforcement; the subsequent investigation and takedown yielded the sanitized backend evidence analyzed in Section~\ref{sec:blackhat-economy}. After disclosure, we observed that at least 148 affected sites among the domains identified by OCL Detector had deployed recommended defenses, including safer MSSO implementations such as masked-number completion.
}

\section{Generative AI Usage}

We used generative AI tools (e.g., large language models) only for minor editorial assistance, such as grammar checking and light style polishing of draft text. All technical content, experimental design, data analysis, and conclusions were created and validated by the authors, and all AI-assisted text was manually reviewed and, when necessary, rewritten by the authors.

\newpage

\section{MSSO Provider}
\label{apx:msso_provider}

Table~\ref{tab:ecosystem_providers} summarizes the MSSO provider landscape identified in our preliminary study. We classify providers by whether they possess the authoritative capability to perform network-level identity verification. MNOs, such as China Mobile, China Telecom, and China Unicom, are direct providers because they control the mapping between a user's active mobile data session and the verified subscriber identity. Resellers, such as Alibaba Cloud and Netease Yidun, do not possess this identity authority; instead, they purchase authentication capabilities from multiple MNOs and rebundle them into unified SDKs or APIs for SPs.

The table also lists the operational artifacts used by OCL Detector. The \texttt{Bundled MNOs} column shows whether a provider exposes a single-MNO service or abstracts over CM, CT, and CU simultaneously, while the \texttt{Seed Endpoint(s)} capture the domains and URL patterns used to recognize MSSO traffic during measurement. Direct MNO endpoints expose the authentication anchor behind a deployment, whereas Reseller endpoints introduce an additional mediation layer in which the downstream SP can be hidden behind the Reseller's identity and \textit{developer credentials}. The same table therefore supports both the seed-endpoint construction in Section~\ref{sec:preliminary_study} and the origin-validation analysis in Section~\ref{sec:prevalence_of_defect_3}.



\begin{table*}[htb]
\centering
\caption{Key MSSO providers identified in our study, categorized by architectural role, service model, and seed endpoints used for detection.}
\label{tab:ecosystem_providers}
\begin{tabular}{@{}llccl@{}}
\toprule
\textbf{Provider} & \textbf{Role} & \textbf{Bundled MNOs} & \textbf{Prevalence (Sites)} & \textbf{Seed Endpoint(s)} \\
\midrule
China Mobile (CM) & MNO & CM & 595 & \makecell[l]{h5auth.cmpassport.com \\ verify.cmpassport.com \\ www.cmpassport.com \\ authsafe.cmpassport.com} \\
China Telecom (CT) & MNO & CT & 380 & \makecell[l]{id6.me/auth \\ e.189.cn \\ open.e.189.cn} \\
China Unicom (CU) & MNO & CU & 237 & \makecell[l]{wosms.cn \\ auth.wosms.cn} \\
\midrule
Alibaba Cloud & Reseller & CM, CT, CU & 79 & \makecell[l]{dypns.aliysm.cn \\ dypns-operator.oss-cn-hangzhou.aliyuncs.com} \\
Baidu & Reseller & CM, CT, CU & 33 & --- \\
Cloopen & Reseller & CM, CT, CU & 3 & aim-mobileauth.yuntongxun.com \\
Geetest & Reseller & CM, CT, CU & 26 & --- \\
Getui & Reseller & CM, CT, CU & 13 & gysdk-min.js$\dagger$ \\
Netease Yidun & Reseller & CM, CT, CU & 17 & \makecell[l]{yidunfe.nosdn.127.net \\ ye.dun.163.com} \\
Qiniu Cloud & Reseller & CM, CT, CU & 15 & --- \\
Shanyan & Reseller & CM, CT, CU & 7 & chuanglan/shanyan$\dagger$ \\
Shumei Technology & Reseller & CM, CT, CU & 2 & api-oneclick-bj.fengkongcloud.com \\
Umeng+ & Reseller & CM, CT, CU & 27 & \makecell[l]{havana-nlogin$\dagger$ \\ g.alicdn.com/jssdk/verify-h5-ui} \\
Jiguang & Reseller & CM, CT, CU & 10 & jverification.jiguang.cn \\
\bottomrule
\end{tabular}
\parbox{\textwidth}{\small \textit{Note:} $\dagger$~URL substring matching patterns for SDK identification on shared CDNs or customer-hosted resources, not independent API domains.}
\end{table*}

\section{Feature Definitions for Script Profiling}
\label{sec:appendix_features}

Table~\ref{tab:all-features} lists the features used by the Abuse Detector to profile JavaScript files captured during dynamic MSSO verification. Since direct observation of user-data exfiltration is unreliable, the detector focuses on the reusable distribution layer: scripts that are repeatedly loaded by downstream sites and that participate in MSSO-related requests. The feature design therefore combines three views of each script: its intrinsic code properties, its deployment context, and its runtime relationship to MSSO traffic.

\hjs{The three feature groups serve different roles in clustering. General static features describe code properties such as entropy, encoded strings, dynamic execution, and event handling. Provider features characterize deployment context, including prevalence and first-party loading. Specific behavioral features connect a script to MSSO execution by measuring whether, and how frequently, it triggers service-signature rules. Their combination supports density-based grouping and anomalous-candidate selection; behavioral interpretation and abuse attribution rely on the subsequent feature analysis, manual review, upstream-provider analysis, and case studies.}

\begin{table*}[hb!]
\centering
\footnotesize
\caption{The canonical feature set used for script clustering: 24 scalar features and two variable-dimensional feature blocks.}
\label{tab:all-features}
\resizebox{\textwidth}{!}{%
\begin{tabular}{@{}lll@{}}
\toprule
\textbf{Category} & \textbf{Feature Name} & \textbf{Definition} \\ \midrule
\textbf{A. General Static Features} & \texttt{script\_size} & Script content length after UTF-8 decoding. \\
 & \texttt{content\_entropy} & Shannon entropy of the script content. \\
 & \texttt{avg\_line\_length} & Script length divided by the number of lines. \\
 & \texttt{string\_ratio} & Fraction of script characters contained in string literals. \\
 & \texttt{long\_string\_num} & Number of string literals with more than 40 content characters. \\
 & \texttt{suspicious\_string\_num} & Number of string literals containing selected suspicious terms. \\
 & \texttt{encoded\_string\_num} & Number of string literals containing hexadecimal or Unicode escapes. \\
 & \texttt{handler\_num} & Number of selected event-handler occurrences. \\
 & \texttt{uses\_eval} & Binary indicator for \texttt{eval()} or \texttt{new Function()}. \\
 & \texttt{uses\_settimeout} & Binary indicator for \texttt{setTimeout}. \\
 & \texttt{uses\_setinterval} & Binary indicator for \texttt{setInterval}. \\
 & \texttt{uses\_navigator\_useragent} & Binary indicator for \texttt{navigator.userAgent}. \\ \midrule
\textbf{B. Provider Features} & \texttt{first\_party\_ratio} & Fraction of observations loaded as a first-party resource. \\
 & \texttt{prevalence\_sites} & Number of unique sites on which the script hash appears. \\
 & \texttt{unique\_source\_domains} & Number of unique source domains. \\
 & \texttt{unique\_paths} & Number of unique observed URL paths. \\
 & \texttt{url\_spec\_keywords} & Number of selected authentication-related keywords present in observed URLs. \\
 & \texttt{unique\_query\_key\_sets} & Number of unique query-key combinations. \\
 & \texttt{path\_is\_hashed} & Binary indicator for selected hashed-build path patterns. \\
 & \texttt{path\_contains\_version} & Binary indicator for selected version path patterns. \\
 & \texttt{has\_id\_param} & Binary indicator for selected identifier query keys. \\ \midrule
\textbf{C. Behavioral Features} & \texttt{rule\_trigger\_count\_log} & $\log(1+n)$ for observed MSSO-rule triggers. \\
 & \texttt{rule\_trigger\_ratio} & Trigger count divided by prevalence sites. \\
 & \texttt{triggering\_site\_ratio} & Unique triggering sites divided by prevalence sites. \\ \midrule
\textbf{D. Feature Blocks} & \texttt{domain\_multi\_hot} & Binary multi-hot encoding of observed source domains. \\
 & \texttt{path\_tfidf} & Mean TF--IDF vector of path-segment unigrams and bigrams. \\ \bottomrule
\end{tabular}%
}
\end{table*}

\section{Masked-Number Completion Figure}
\label{apx:masked_number_completion}

\begin{figure}[t]
    \centering
    \includegraphics[width=.9\linewidth]{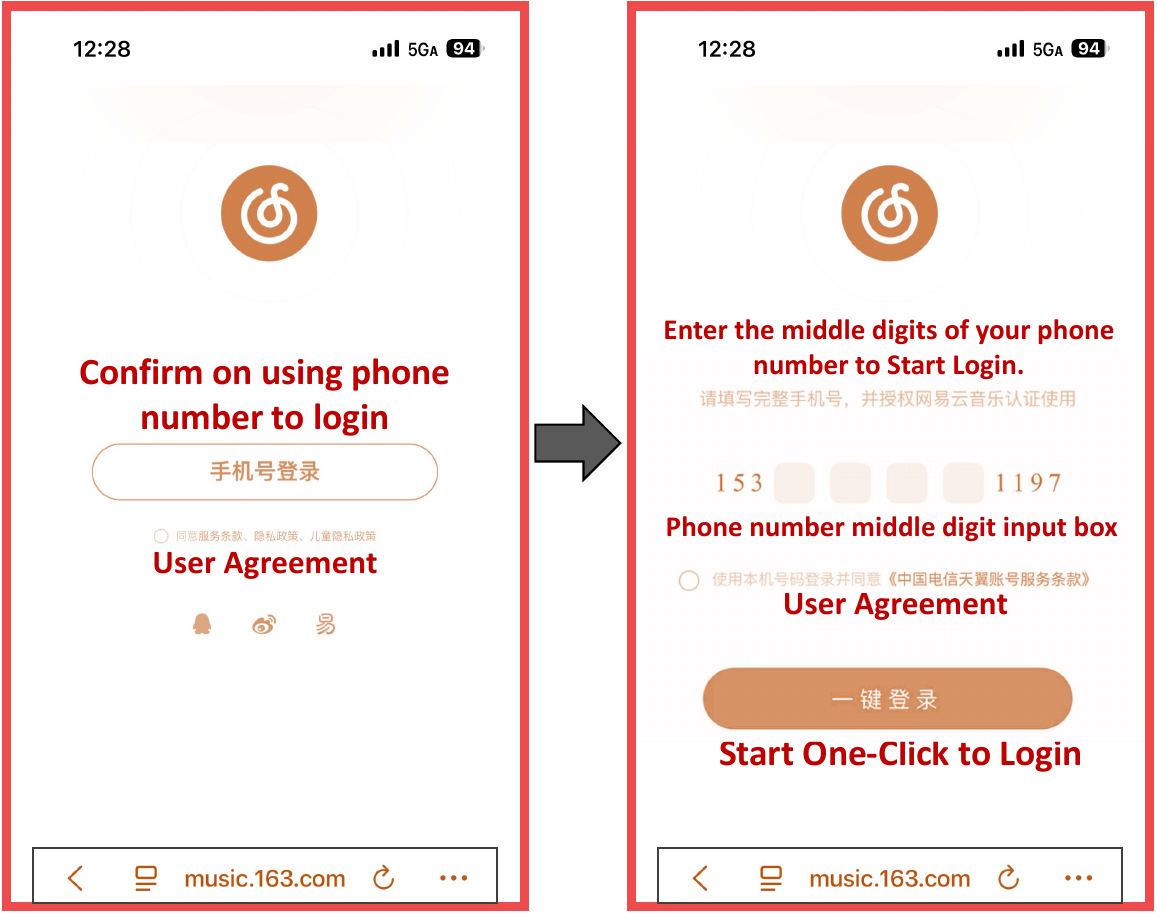}
    \caption{An example of masked-number completion verification in web-based MSSO.}
    \label{fig:masked-number-completion}
\end{figure}

Figure~\ref{fig:masked-number-completion} shows the masked-number completion mode discussed in Section~\ref{sec:discussion}. The key change is that the pre-authentication phase no longer exposes enough information for an automated script to complete the flow alone. After receiving a partially masked phone number, the user must fill in the hidden digits before the MNO issues a User Identity Token.

This interaction preserves the central usability goal of MSSO: the user still authenticates through a lightweight phone-number confirmation rather than a password or SMS code. At the same time, it restores a user-participation signal that the conventional one-tap flow lacks. An adversary who only controls a webpage script can trigger background requests, but cannot infer the missing digits without either user input or brute force; MNO-side rate limiting can further make such brute force impractical.


\section{Consent-Bypass Figure}

\begin{figure*}[t]
    \centering
    \includegraphics[width=.9\linewidth]{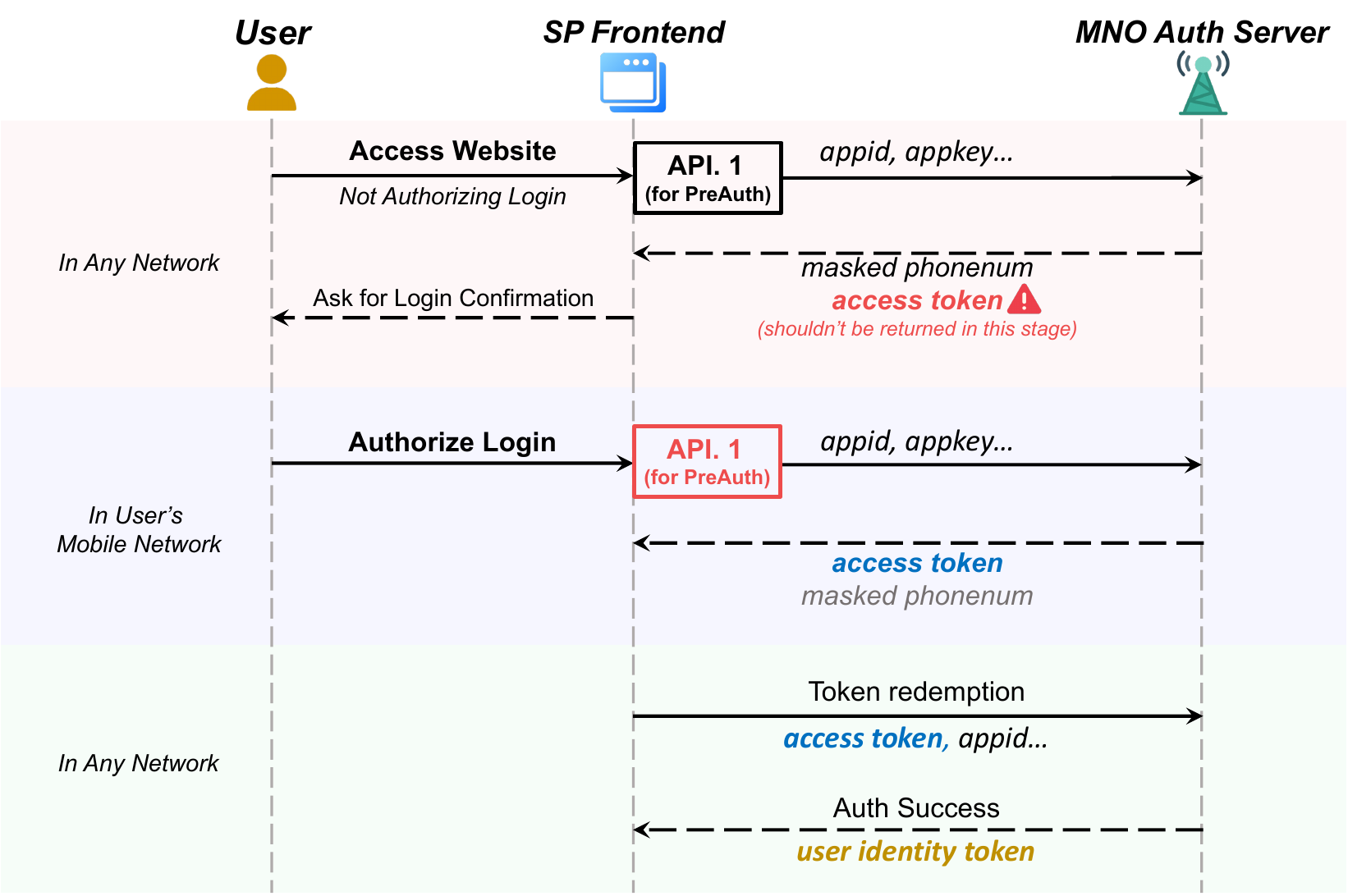}
    \caption{The workflow of consent-bypass flaw}
    \label{fig:consent-bypass-flaw}
\end{figure*}

Figure~\ref{fig:consent-bypass-flaw} shows how \textit{Trust Defect 1 (Unenforced User Consent)} manifests as consent bypass in Section~\ref{sec:prevalence_of_defect_1}. The defect collapses the intended separation between pre-authentication and token issuance. In the intended design, pre-authentication is a low-privilege preparation step: it returns only display data, such as a masked phone number, so the user can decide whether to continue. The high-privilege tokens (e.g., the access token and the resulting User Identity Token) should be generated only after the user explicitly confirms the login action.

The defective workflow moves this high-privilege token issuance into the pre-authentication path. As a result, a request that appears to be a harmless page-load optimization can already return identity-bearing material before any user click. This is the concrete mechanism behind Trust Defect 1: the MNO cannot verify that user consent occurred, and an adversary-controlled page or script can obtain a token simply by invoking the pre-fetch API.

\end{document}